\documentclass[preprint,aps,prc,showpacs,nofootinbib,secnumarabic]{revtex4-1}

\usepackage{graphicx}% Include figure files
\usepackage{bm}% bold math
\usepackage{color}

\newcommand{\ltsim}{\protect\raisebox{-0.5ex}{$\:\stackrel{\textstyle <}
	{\sim}\:$}}
\newcommand{\gtsim}{\protect\raisebox{-0.5ex}{$\:\stackrel{\textstyle >}
	{\sim}\:$}}

\begin{document}

\title{Color transparency in $\pi^-$-induced dilepton production on nuclei}

\author{A.B. Larionov$^{1,2}$\footnote{Present address: Institut f\"ur Theoretische Physik, Universit\"at Giessen,
             D-35392 Giessen, Germany}, M. Strikman$^3$, M. Bleicher$^{1,4}$}

\affiliation{$^1$Frankfurt Institute for Advanced Studies (FIAS), 
             D-60438 Frankfurt am Main, Germany\\ 
             $^2$National Research Center "Kurchatov Institute", 
             123182 Moscow, Russia\\
             $^3$Pennsylvania State University, University Park, PA 16802, USA\\
             $^4$Institut f\"ur Theoretische Physik, J.W. Goethe-Universit\"at,
             D-60438 Frankfurt am Main, Germany}

\date{\today}

\begin{abstract}
We  argue that the observation of the color transparency effect in the semiexclusive $A(\pi^-,l^+ l^-)$ process
is important for determining whether it is possible to extract the generalized parton distributions of the nucleon from 
the elementary reaction  $\pi^- p \to l^+ l^- n$ at $p_{\rm lab}=15-20$ GeV/c at small $|t|$ and large invariant mass of the
dilepton pair $l^+ l^-$.
Assuming that the transverse size of the pionic $q \bar q$ pair in the hard interaction point is similar to the one 
in the reaction $\gamma^* p\to \pi^+ n$ studied at JLab we predict large color transparency effects in the discussed kinematic range.
We also suggest that the semiexclusive $\rho^0$ production in $\pi^-$-induced reactions 
in the same beam momentum region may provide new information on the dynamics of the interaction in the non-vacuum channel,
while the $J/\psi$ production can be used to get information on $J/\psi N$ total interaction cross section.  
\end{abstract}

% Factorization

\maketitle

\section{Introduction}

It was suggested in \cite{Brodsky,Mueller} to test the hypothesis that small size  quark - antiquark and 
three quark configurations dominate in large angle two body elastic scattering by studying the $A$-dependence 
of quasielastic scattering off nuclei at large $|t|$ and $|u|$. 
If this is the case, the color transparency (CT) phenomenon (the suppression of the interaction of small color dipoles)
would lead to a complete disappearance of nuclear shadowing. 
For the review of CT and other color-coherent phenomena we refer the reader to Refs. \cite{Frankfurt:1994hf,Jain:1995dd,Dutta:2012ii}.

Later on the factorization theorem was derived for the exclusive meson production by longitudinally polarized photons 
\cite{Brodsky:1994kf,Collins:1996fb}. 
The proof essentially relied on involving the CT feature of  QCD.

CT has been experimentally observed in several high-energy process:
pion dissociation into dijets \cite{Aitala:2000hc} 
predicted theoretically in Ref. \cite{Frankfurt:1993it},
incoherent photoproduction of $J/\psi$'s 
\cite{Sokoloff:1986bu}, and exclusive electroproduction of $\rho^0$'s and $J/\psi$'s at HERA 
\cite{Chekanov:2004mw,Chekanov:2007zr}. At intermediate energies, $E_{\rm beam} \sim 10$ GeV, 
however, the CT is blurred by the expansion of the point-like configurations
towards normal hadronic size. There are indications for the onset of CT in
experiments on $\pi^+$ \cite{Clasie:2007aa} and $\rho^0$ \cite{ElFassi:2012nr} electroproduction
at intermediate energies at JLab.
However, no evidence for CT was reported so far in the process $A(e,e^\prime p)(A-1)^*$ 
indicating that squeezing a nucleon requires larger $Q^2$ than in the meson case.
Here and below the word ``squeezing'' means the preferential selection of the color singlet, small transverse size
configurations in exclusive processes on a nucleus at high momentum transfer (cf. \cite{Dutta:2012ii})
or at high time-like momentum of the intermediate $\gamma^*$.

A very strong CT signal has been predicted for the pionic quasielastic knockout of protons from nuclei, i.e.
for the $A(\pi,\pi p)$ reaction at $p_{\rm lab}=200$ GeV/c for $|t| \sim 2-10$ GeV$^2$ \cite{Miller:2010eh}.
Significant CT signal is expected in $A(\gamma^*,\rho p)$ reaction at the virtual photon energies
$\nu \geq 10$ GeV \cite{Howell:2013ikq} requiring measurements at fixed photon coherence length.
These large $|t|$ semiexclusive reactions include the CT effects for 2 or 3 particles.
In general, this complicates their theoretical description.
     
In almost all existing measurements (except for pion diffractive dissociation \cite{Aitala:2000hc}),
the CT was observed or searched for in the interactions of the {\it outgoing} hadrons with the nuclear
target residue. On the other hand, the nuclear transparency due to the absorption of the {\it incoming}
hadron might even be a cleaner observable, since the momentum and the type of the particle are under
the full control. 

The experimental opportunities for these studies exist at the high-momentum 
beam line at J-PARC which will provide a secondary $\pi^-$ beam of momenta up to $p_{\rm lab}=20$ GeV/c.
A program for the studies of the Drell-Yan process with  secondary $\pi^-$ beams 
is under discussion at J-PARC (cf. \cite{Kumano:2015gna} and Refs. therein). 
It includes the study of the exclusive reaction
$\pi^- p\to l^+ l^- n$ which is complementary to the DIS process $\gamma^* p\to \pi^+ n$. 
For the meson electroproduction in the case of longitudinally polarized photons the factorization theorem 
has been proven \cite{Collins:1996fb}. 
This opens the possibility to measure the generalized parton distributions (GPD's) in the nucleon at large $Q^2$. 

So far no similar factorization theorem was proven for the exclusive Drell-Yan process.
Nevertheless, factorization was assumed in %a number of papers 
Refs. \cite{Berger:2001zn,Goloskokov:2015zsa}
where the amplitude of this process is given by an expression similar to the leading twist expression  
valid for the pion electroproduction \cite{Mankiewicz:1998kg,Frankfurt:1999fp}.

In particular, it has been suggested in Ref. \cite{Berger:2001zn} that the nucleon GPD's can be probed
in exclusive reactions $\pi^- p \to l^+ l^- n$ and $\pi^+ n \to l^+ l^- p$ with large invariant mass
of the dilepton pair, $M_{l^+l^-} \sim 2-3$ GeV, and small squared momentum transfer to the nucleon,
$|t| < 0.4$ GeV$^2$. In the leading order in the factorization scale, $M_{l^+l^-}$, 
and leading order in $\alpha_s$ the amplitides of the processes $\pi^- p \to \gamma^* n$ 
and $\pi^+ n \to \gamma^* p$ with time-like photon can be related, respectively, to the amplitudes 
of the processes $\gamma^* p \to \pi^+ n$ and $\gamma^* n \to \pi^- p$ with space-like photon. 
The concrete calculations have been performed 
in Ref. \cite{Berger:2001zn} by using ansatz $\tilde H^u - \tilde H^d \propto g_A(t)$ where
$g_A(t)$ is the axial form factor of the nucleon and a form motivated by the chiral soliton model of the
nucleon, $\tilde E^u - \tilde E^d \propto (m_\pi^2-t)^{-1}$. This resulted in the differential cross section,
$d\sigma/dM_{l^+l^-}^2dt~(\pi^- p \to \gamma^* n)$, for longitudinal $\gamma^*$ of the order of 1 pb/GeV$^4$ 
with large uncertainties due to the input GPD's. 
In the recent work \cite{Goloskokov:2015zsa} the extended calculations of the $\pi^- p \to l^+ l^- n$
process have been performed. In contrast to Ref. \cite{Berger:2001zn}, the authors of Ref. \cite{Goloskokov:2015zsa}
have taken into account the transversity GPD's, $H_T$ and $\tilde E_T$, and used the phenomenological one-pion exchange term
for treating the pion pole contribution. The longitudinal cross sections predicted in Ref. \cite{Goloskokov:2015zsa}
are about 40 times bigger then those of Ref. \cite{Berger:2001zn}, mostly due to the different treatment of the pion
pole. We would like to note that the presence of CT in such models is closely related 
to the issue of shrinking of $q \bar q$ configurations in the pion
contributing to the pion form factor in the time-like region.
Anyway it is clear that the availability of experimental data on the $\pi^- p \to l^+ l^- n$ reaction
would allow to constrain the $p \to n$ transition GPD's provided that the factorization works in this case.

Overall the issue of factorization in the $\pi^- p \to l^+ l^- n$ reaction remains a matter of
debate. In Ref. \cite{Muller:2012yq} authors give arguments based on the comparison of  the
analytic structure of the discussed reaction and the corresponding DIS exclusive
reaction that  the factorization holds at large $Q^2$.
On the other hand, J.W.~Qiu has suggested \cite{Qiu_J-PARC_2015} that factorization may be violated
for the exclusive Drell-Yan process.

The presence of CT in a hard process is a necessary condition for the validity of factorization and the possibility 
to extract GPD's from the exclusive process. 
Indeed, the factorization of the elementary reaction amplitude to the soft part described by GPD's and meson distribution
amplitudes and hard part described by perturbative QCD is only possible if the multiple soft gluonic interactions are suppressed.
However, this may only take place if the participating $q \bar q$ and $qqq$ configurations are color neutral and have
transverse size substantially smaller than ``normal'' hadrons such that gluonic fields coupled to the different quarks almost cancel each other.     
In the case of nuclear target, such a cancellation unavoidably results in the CT phenomenon. 
 
Hence, we propose to explore the $A$-dependence of the exclusive Drell-Yan process to find out 
whether squeezing does indeed occur for this process.  
We will theoretically explore the magnitude of the CT effects in the semiexclusive process $A(\pi^-,l^+l^-)$ 
at $p_{\rm lab} \sim $ few tens of GeV/c due to the $l^+l^-$ production on the nuclear target proton, $\pi^- p \to l^+ l^- n$. 
In our numerical studies we will assume that squeezing at positive and negative $Q^2$ is similar.

The paper is organized as following.
In Section \ref{model} we describe the model starting from the Glauber approximation and then extending it to account
for the CT effects. Section \ref{kinem} contains the estimates for the lepton momentum resolution necessary to separate 
the final states where a slow nucleon and a slow $\Delta$ isobar are produced in the processes 
$\pi^- p \to l^+l^- B$, $B=n,\Delta^0$.
Section \ref{Transparency} contains the results of the numerical calculations of the nuclear transparency ratios for 
a variety of the processes. As a first benchmark, we consider the pion electroproduction 
$A(e,e^\prime \pi^+)$ in the region where JLab data \cite{Clasie:2007aa} are available. 
Then we discuss in detail the possible CT effect
for the $A(\pi^-,l^+l^-)$ process which is the main focus of this work. Finally, the vector meson production $A(\pi^-,V)$,
$V=\rho^0, J/\psi$ is addressed. In the concluding Section \ref{concl} we discuss the main results of this study.
  
\section{Model}
\label{model}

\begin{figure}
\includegraphics[scale = 0.6]{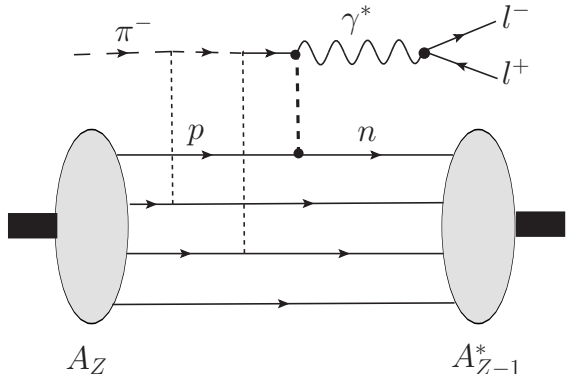}
\caption{\label{fig:piA2ee} The Feynman graph describing the amplitude of the semiexclusive
$A_Z(\pi^-,l^+l^-)A^*_{Z-1}$ process. Thin short-dashed lines represent the pion-nucleon soft 
elastic scattering amplitudes, while a thick short-dashed line represents the transition
amplitude $\pi^- p \to \gamma^* n$. The gray ellipsoids correspond to the wave functions 
of the initial ground state nucleus $A_Z$ and final excited nucleus $A^*_{Z-1}$.}
\end{figure}
The process of exclusive $l^+l^-$-pair production in the interaction of a $\pi^-$-meson with a 
nucleus is shown in Fig.~\ref{fig:piA2ee}. We will assume that the momentum transfer in the $p \to n$ transition 
is small, i.e. of the order of a few hundred MeV/c. 
(See Fig.~\ref{fig:kz_vs_m2ee} and discussion in Sec.~\ref{kinem} below.)
Thus, the outgoing nucleon system can be viewed as an excited state of the $A_{Z-1}$ nucleus. 
The amplitude of Fig.~\ref{fig:piA2ee} can be calculated within the
generalized eikonal approximation \cite{Frankfurt:1996xx,Sargsian:2001ax,Larionov:2013nga}. 
By keeping only absorptive terms in the expansion of the matrix element squared (i.e. neglecting the terms, where
soft pion rescattering occurs on the same nucleon in the direct and conjugated amplitudes, 
which is a good approximation for small transverse momenta of the dilepton pair) the differential cross section 
of the dilepton production on the nucleus can be expressed as
\begin{equation}
    \frac{d^4\sigma_{\pi^- A \to l^-l^+}}{d^4q} 
   =  v_{\pi}^{-1} \int d^3r {\cal P}_{\pi,{\rm surv}}(\mathbf{b},-\infty,z)
      \int \frac{d^3p}{(2\pi)^3} v_{\pi p} \frac{d^4\sigma_{\pi^- p \to l^-l^+ n}}{d^4q} 
      \sum_{j=1}^Z f_j(\mathbf{r},\mathbf{p})~,      \label{dsigdMd3q}
\end{equation}
where $q=p_{l^-}+p_{l^+}-p_{\pi}$ is the four-momentum transfer to the dilepton pair,  
$v_{\pi}=p_{\rm lab}/E_{\pi}$ is the pion velocity in the laboratory system, 
$E_{\pi}=\sqrt{m_{\pi}^2+p_{\rm lab}^2}$ is the pion energy,
$v_{\pi p}=\sqrt{(p_{\pi}p)^2-m_{\pi}^2m_p^2}/E_{\pi}E_p$ is the relative velocity
of the pion and the target proton, and 
\begin{equation}
   f_j(\mathbf{r},\mathbf{p})=\int d^3r^\prime   \phi_j^*(\mathbf{r}+\mathbf{r}^\prime/2) 
                                               \phi_j(\mathbf{r}-\mathbf{r}^\prime/2)
                                               {\rm e}^{i \mathbf{p} \mathbf{r}^\prime}   \label{f_j}
\end{equation}
is the Wigner density of the occupied proton state $j$ including orbital and spin quantum numbers.
We neglect spin-flip transitions. Thus, in Eq.(\ref{f_j}), the trace is assumed over spin variables 
of the single-particle wave functions.
In Eq.(\ref{dsigdMd3q}), the depletion of the pion flux is accounted for by the pion survival probability
\begin{equation}
    {\cal P}_{\pi,{\rm surv}}(\mathbf{b},-\infty,z) 
   = \exp\left(-\sigma_{\pi N} \int\limits_{-\infty}^zdz^\prime \rho(\mathbf{b},z^\prime)\right)~,   \label{Psurv}
\end{equation}
where $\sigma_{\pi N}$ is the total pion-nucleon cross section and $\rho(\mathbf{b},z^\prime)$
is the nucleon density. Equation (\ref{dsigdMd3q}) is derived by summing over a complete set of the
wave functions of the final nuclear system $A^*_{Z-1}$. (The Pauli blocking 
for the outgoing neutron is automatically included as the wave functions of the complete 
set are antisymmetric.) 
In other words, we used the closure assumption without any restriction on the momentum transfer.

In the simplest approximation the Fermi motion can be neglected 
\footnote{Hence we neglect corrections related to the expected  rather
moderate dependence of the elementary cross section on the incident energy:
$d \sigma/d t  \propto s^{-n}$ where $n=2~(1)$ assuming the
dominance of the $\pi$($\rho$)-meson Regge trajectory. 
This enhances the contribution of the scattering off the protons moving parallel to $\mathbf{p}_\pi$ 
as compared to the ones moving antiparallel.}.
Thus, the proton momentum in the relative velocity $v_{\pi p}$ and
in the elementary cross section $d^4\sigma_{\pi^- p \to l^-l^+ n}/d^4q$ can be set to zero, i.e the target
proton is assumed to be quasifree. This leads us to the classical formula for the nuclear transparency
ratio
\begin{equation}
    T_{l^-l^+} =\frac{d^4\sigma_{\pi^- A \to l^-l^+} /d^4q}{Zd^4\sigma_{\pi^- p \to l^-l^+ n}/ d^4q}
     = \frac{1}{Z} \int d^3r {\cal P}_{\pi,{\rm surv}}(\mathbf{b},-\infty,z) \rho_p(\mathbf{r})~,    \label{T_e+e-}
\end{equation}
where
\begin{equation}
   \rho_p(\mathbf{r}) = \int \frac{d^3p}{(2\pi)^3} \sum_{j=1}^Z f_j(\mathbf{r},\mathbf{p})    \label{rho_p}
\end{equation}
is the proton density.

Until this point we discussed the classical Glauber picture which does not include any features of the quark structure of the
pion. In factorization approaches, the internal structure of a meson is described by its light-cone wave function 
encoding the QCD dynamics. The general proof of factorization for the electroproduction of mesons 
by the longitudinally polarized photons \cite{Collins:1996fb} is valid for production of any mesons and as a part of the proof
the dominance of small size configurations in the discussed process was demonstrated. 
Thus, CT should also be present for this class of processes.  
As we mentioned above so far no proof of factorization was put forward for the case of 
dilepton production in meson-induced reactions.
However, there exist experimental indications that squeezing is a more general phenomenon. 
For example, squeezing seems to be present for the production of $\rho$ mesons by transversely polarized photons.
So for our estimates we will  assume  that the basic mechanism of CT is similar to the case of meson electroproduction, 
though in principle there maybe a difference between squeezing in the case of longitudinally and transversely polarized photon. 
The momentum scale of $M^2_{l^+l^-}$ governs the degree of the shrinkage of the transverse size of the meson while it approaches 
the interacting proton of the target. Such a transversely squeezed pion (which can be considered as a small color $q \bar q$ dipole)
interacts  with nucleon with an effectively reduced 
cross section that can be evaluated within the quantum diffusion model \cite{Farrar:1988me} as follows:
\begin{equation}
   \sigma_{\pi N}^{\rm eff}(p_{\pi},z)
  =\sigma_{\pi N}(p_{\pi}) \left(\left[ \frac{z}{l_{\pi}}
    + \frac{\langle n^2k_t^2\rangle}{M^2_{l^+l^-}} \left(1-\frac{z}{l_{\pi}}\right) \right]
    \Theta(l_{\pi}-z) +\Theta(z-l_{\pi})\right)~.         \label{sigma_piN_eff}
\end{equation}
Here $n=2$ is the number of valence (anti)quarks, $\langle k_t^2 \rangle^{1/2} \simeq 0.35$ GeV/c is the average transverse momentum of a quark in a hadron,
and
\begin{equation}
    l_{\pi}=\frac{2p_{\rm lab}}{\Delta M^2}     \label{l_pi}
\end{equation}
is the pion coherence length.  We will use the value $\Delta M^2 = 0.7$ GeV$^2$ in default calculations.
We denote the longitudinal distance to the point where hard interaction occurs as $z$.
Note that Eq.(\ref{sigma_piN_eff}) is applicable not only in the cases when the pion was produced in the hard process, 
but also when the hard process was induced by an incoming pion.
The differential cross section of dilepton production on the nucleus and the transparency can be generalized for the processes with CT
by replacing the pion-nucleon cross section by the effective one in the pion survival probability (\ref{Psurv}) as
\begin{equation}
    {\cal P}_{\pi,{\rm surv}}^{\rm CT}(\mathbf{b},-\infty,z) 
   = \exp\left(-\int\limits_{-\infty}^zdz^\prime \sigma_{\pi N}^{\rm eff}(p_{\pi},z-z^\prime) \rho(\mathbf{b},z^\prime)\right)~.   \label{Psurv^CT}
\end{equation}

\begin{figure}
\begin{center}
\includegraphics[scale = 0.4]{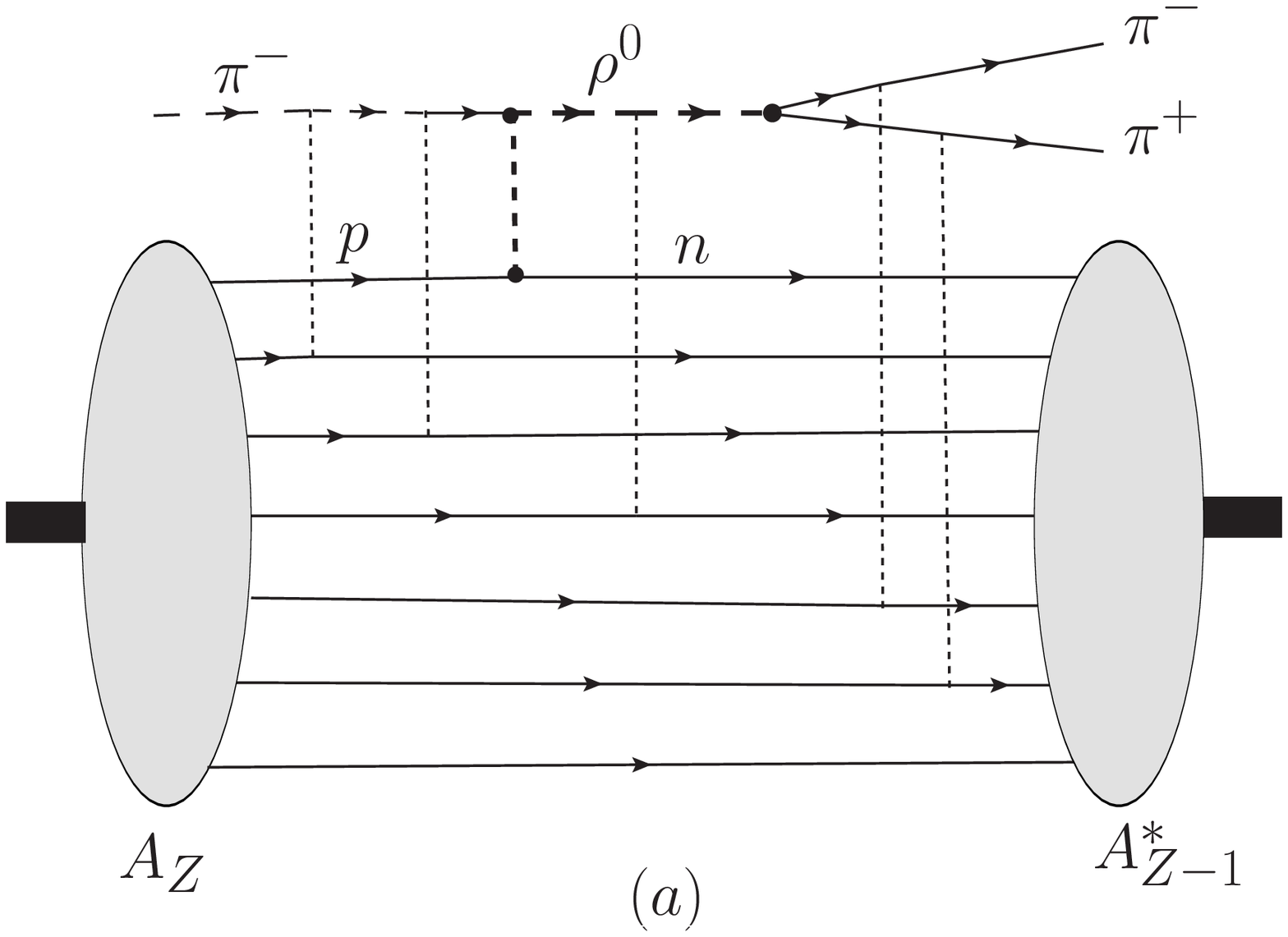}
\includegraphics[scale = 0.4]{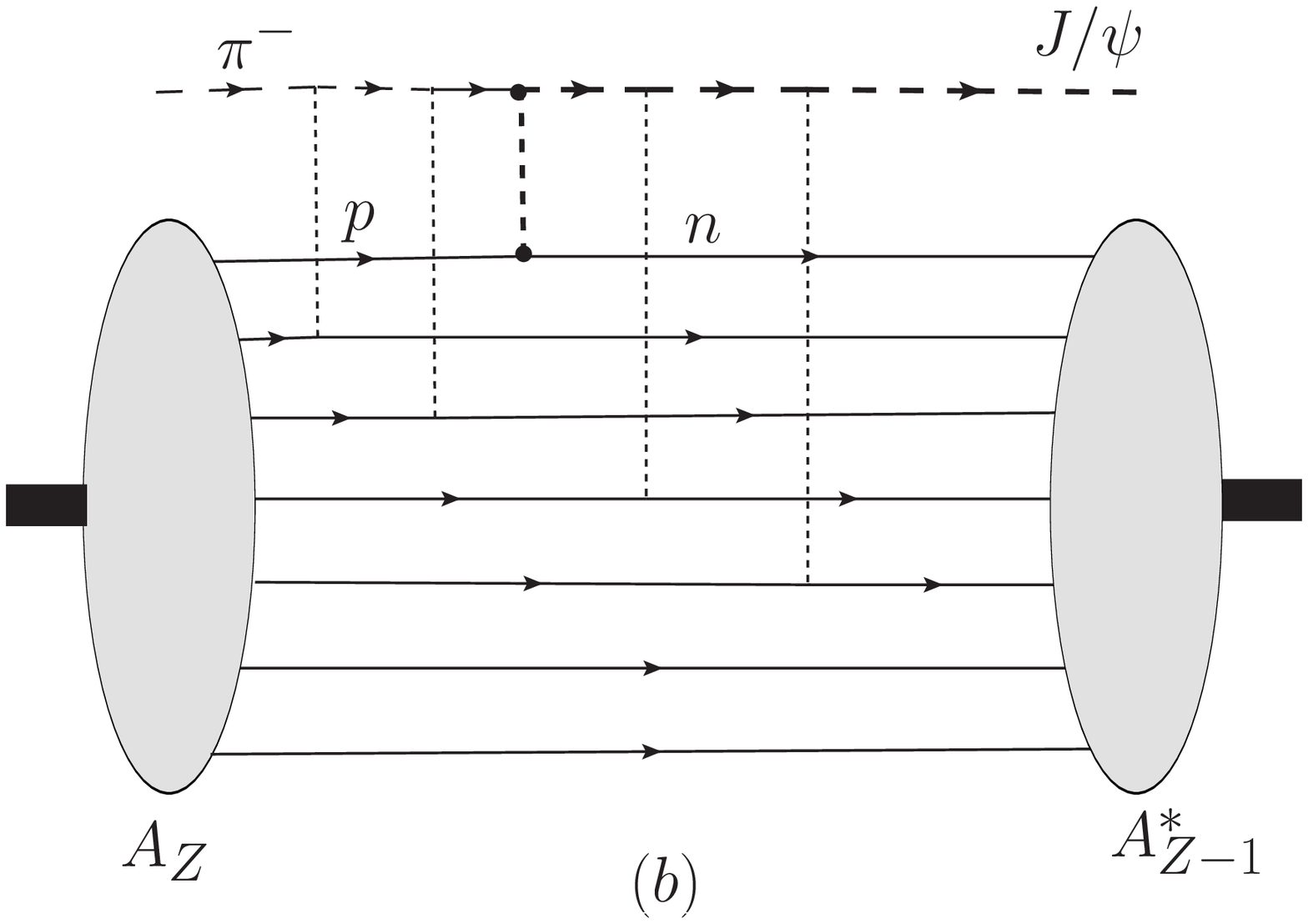}
\end{center}
\caption{\label{fig:piA2vecMes} The Feynman graphs describing the amplitudes of the semiexclusive
$A_Z(\pi^-,\pi^+\pi^-)A^*_{Z-1}$ process mediated by $\rho^0$ (a) and $A_Z(\pi^-,J/\psi)A^*_{Z-1}$ process (b).
Thick short-dashed lines denote $\pi^- p \to V n$ transition amplitudes, $V=\rho^0, J/\psi$.
Other notations are similar to Fig.~\ref{fig:piA2ee}.}
\end{figure}
The process $A_Z(\pi^-,l^+l^-)A^*_{Z-1}$ with large invariant mass of $l^+l^-$ pair represents a clean way to measure
CT for the incoming pion. It is also possible to study other final states in the $\pi^-$-nucleus reactions
with charge exchange in similar kinematical conditions. Two of such channels, $A_Z(\pi^-,\pi^+\pi^-)A^*_{Z-1}$ with 
intermediate $\rho^0$ and $A_Z(\pi^-,J/\psi)A^*_{Z-1}$, are shown in Fig.~\ref{fig:piA2vecMes}.
These two reactions could provide complementary information to the already existing studies of the $\rho$ 
\cite{Hufner:1996dr,Frankfurt:2008pz,Gallmeister:2010wn,ElFassi:2012nr}
and $J/\psi$ \cite{Anderson:1976hi,Sokoloff:1986bu,Farrar:1989vr,Gerschel:1993uh,Kharzeev:1996yx,Hufner:1997jg,Gerland:1998bz,Gerland:2005ca,Larionov:2013axa}
attenuation in the nuclear medium. The corresponding transparencies can be expressed as
\begin{equation}
    T_V = \frac{1}{Z} \int d^3r {\cal P}_{\pi,{\rm surv}}(\mathbf{b},-\infty,z) \rho_p(\mathbf{r})
                                          {\cal P}_{V,{\rm surv}}(\mathbf{b},z,+\infty)~,    \label{T_V}
\end{equation}
with $V=\rho, J/\psi$. The survival probabilities of the vector mesons are expressed as
\begin{equation}
   {\cal P}_{V,{\rm surv}}(\mathbf{b},z,+\infty) 
  = \exp\left(-\int\limits_z^{+\infty}dz^\prime \sigma_{V N}^{\rm eff}(p_V,z^\prime-z) \rho(\mathbf{b},z^\prime)\right)~.   \label{Psurv_V}
\end{equation}
In the case of $J/\psi$ production, since the size of the ${J/\psi}$ is much smaller than for pion
the CT effects are important.
Moreover, the overlap integral in the $\pi^- q\bar q$ vertex may select larger transverse momenta in the $J/\psi$ wave function.
Thus, we evaluate the effective $J/\psi N$ cross section within the quantum diffusion model which gives
\begin{equation}
   \sigma_{J/\psi N}^{\rm eff}(p_{J/\psi},z)
  =\sigma_{J/\psi N} \left(\left[ \frac{z}{l_{J/\psi}}
    + \frac{\langle n^2k_t^2\rangle}{M^2_{J/\psi}} \left(1-\frac{z}{l_{J/\psi}}\right) \right]
    \Theta(l_{J/\psi}-z) +\Theta(z-l_{J/\psi})\right)~,          \label{sigma_JPsiN_eff}
\end{equation}
similar to Eq.(\ref{sigma_piN_eff}) for the $\pi N$ effective cross section. 
Note that in the $J/\psi$ case characteristic $k_t$ are much larger than for the pion - on the scale of $0.8-1~\mbox{GeV}/c$.
This means that the $J/\psi N$ cross section is reduced by approximately a factor of 2 at the hard interaction point.
The interaction cross sections of the $q \bar q$ configurations with the nucleon just before and after the hard interaction 
should be the same. To match this condition in calculations of $J/\psi$ production, we have modified the $\langle k_t^2 \rangle^{1/2}$ value 
in Eq.(\ref{sigma_piN_eff}) for the $\pi N$ effective cross section.
Since the $J/\psi N$ total cross section, $\sigma_{J/\psi N}$, is not well known, we will apply 
the value $\sigma_{J/\psi N}=4$ mb for the default calculations. This value is motivated by the analysis of the $J/\psi$ 
transparencies taking into account the contribution from the radiative decays of $\chi_c$ states \cite{Gerland:1998bz}.
The $J/\psi$ coherence length can be evaluated as (cf. \cite{Farrar:1988me,Gerland:1998bz})
\begin{equation}
   l_{J/\psi}     \simeq  \frac{2p_{J/\psi}}{m_{\psi^\prime}^2-m_{J/\psi}^2}~. \label{l_Jpsi}
\end{equation}
For the energies discussed here $ l_{J/\psi} $ remains small and hence expansion effects are small.
In the case of $\rho$-meson production, the expansion effects are negligible since we are dealing with a soft process.
The survival probability of the $\rho$-meson should, however, be corrected for the $\rho$ decay inside nucleus.
This leads to the following expression for the $\rho N$ effective cross section (cf. \cite{Frankfurt:2008pz}):
\begin{equation}
    \sigma_{\rho N}^{\rm eff}(p_{\rho},z)= \sigma_{\rho N} \exp\left(-\frac{\Gamma_\rho}{\gamma}z\right) 
            + 2 \sigma_{\pi N}(p_{\rho}/2) \left[1-\exp\left(-\frac{\Gamma_\rho}{\gamma}z\right)\right]~,    \label{sigma_rhoN}
\end{equation}
where $\Gamma_\rho=0.149$ GeV is the decay width of the $\rho$ meson at rest and $\gamma=\sqrt{m_\rho^2+p_{\rho}^2}/m_\rho$ 
is the Lorentz factor.
In numerical calculations, for the total pion-nucleon cross section we use the weighted value
\begin{equation}
   \sigma_{\pi N}(p_\pi) = (Z\sigma_{\pi^- p}(p_\pi) + (A-Z)\sigma_{\pi^+ p}(p_\pi))/A~,   \label{sigma_piN}
 \end{equation}
where for the total $\pi^- p$ and $\pi^+ p$ cross sections the PDG parameterization \cite{Eidelman:2004wy} is adopted.
In default calculations, for the total $\rho N$ cross section, $\sigma_{\rho N}$, we assume the constant value of 25 mb. 
It is worth noting here that at high energies the mesons interact in frozen configurations via 
vacuum exchange  with a strength which fluctuates around the average value given by $\sigma_{\rm tot}$.
On the other hand, the reaction $\pi^-p\to \rho^0 n$ is described by the exchange of non-vacuum reggeon in $t$-channel and  
may be dominated by configurations in mesons with sizes larger or smaller than the average. 
For our rough estimates we neglect this effect.

\section{Kinematical constraints}
\label{kinem}

\begin{figure}
\begin{tabular}{ll}
\includegraphics[scale = 0.43]{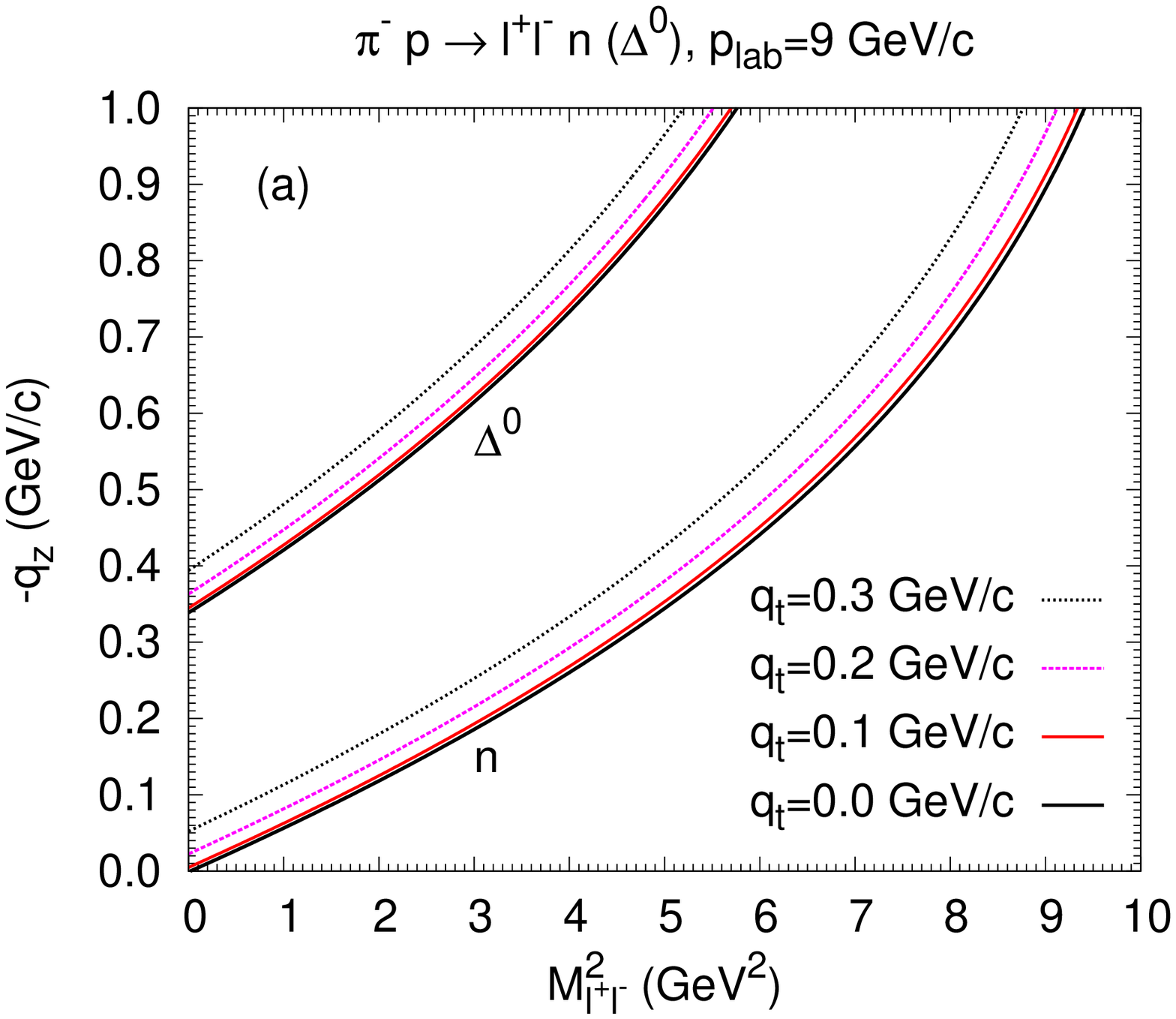} &
\includegraphics[scale = 0.43]{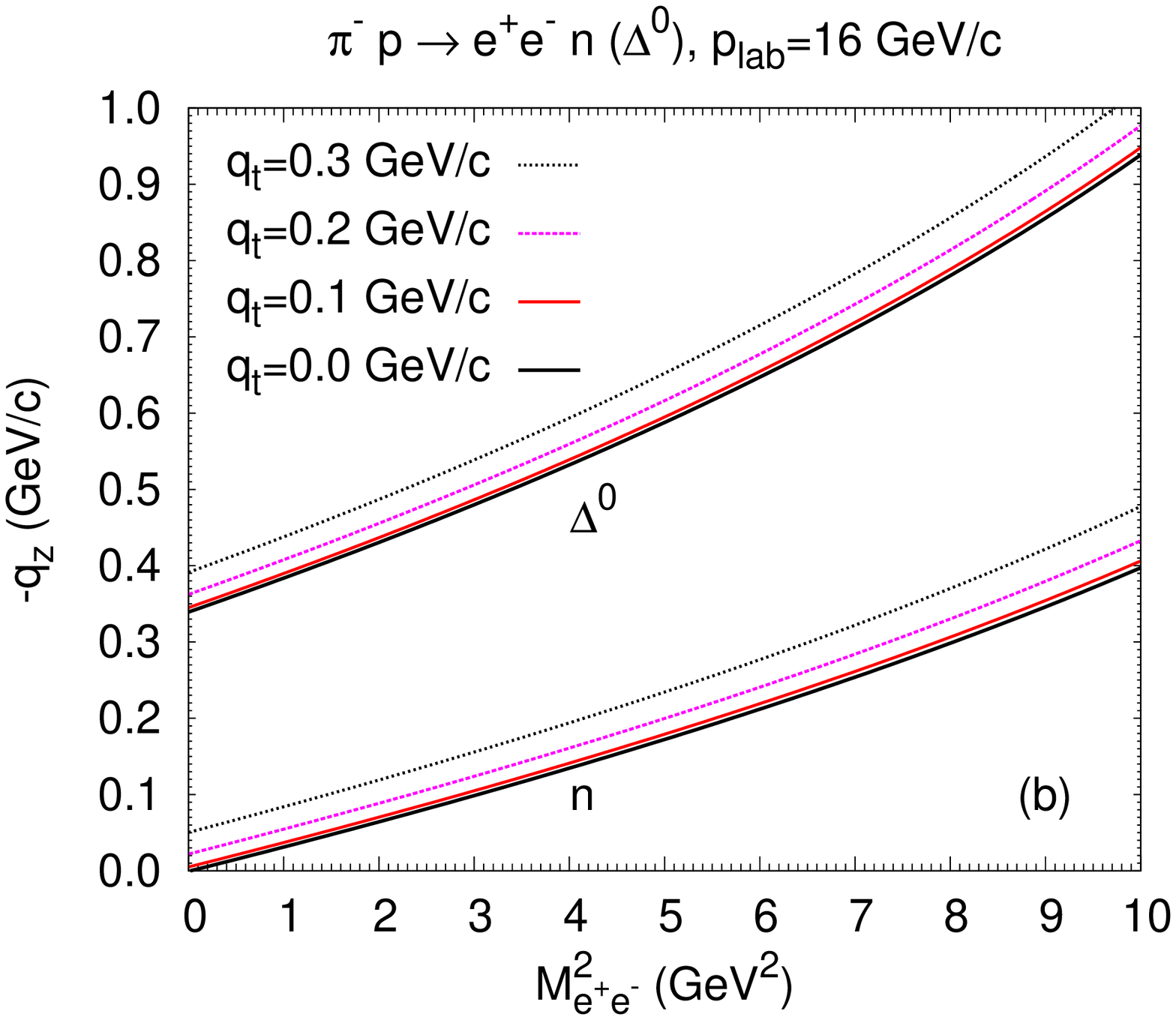}\\
\includegraphics[scale = 0.43]{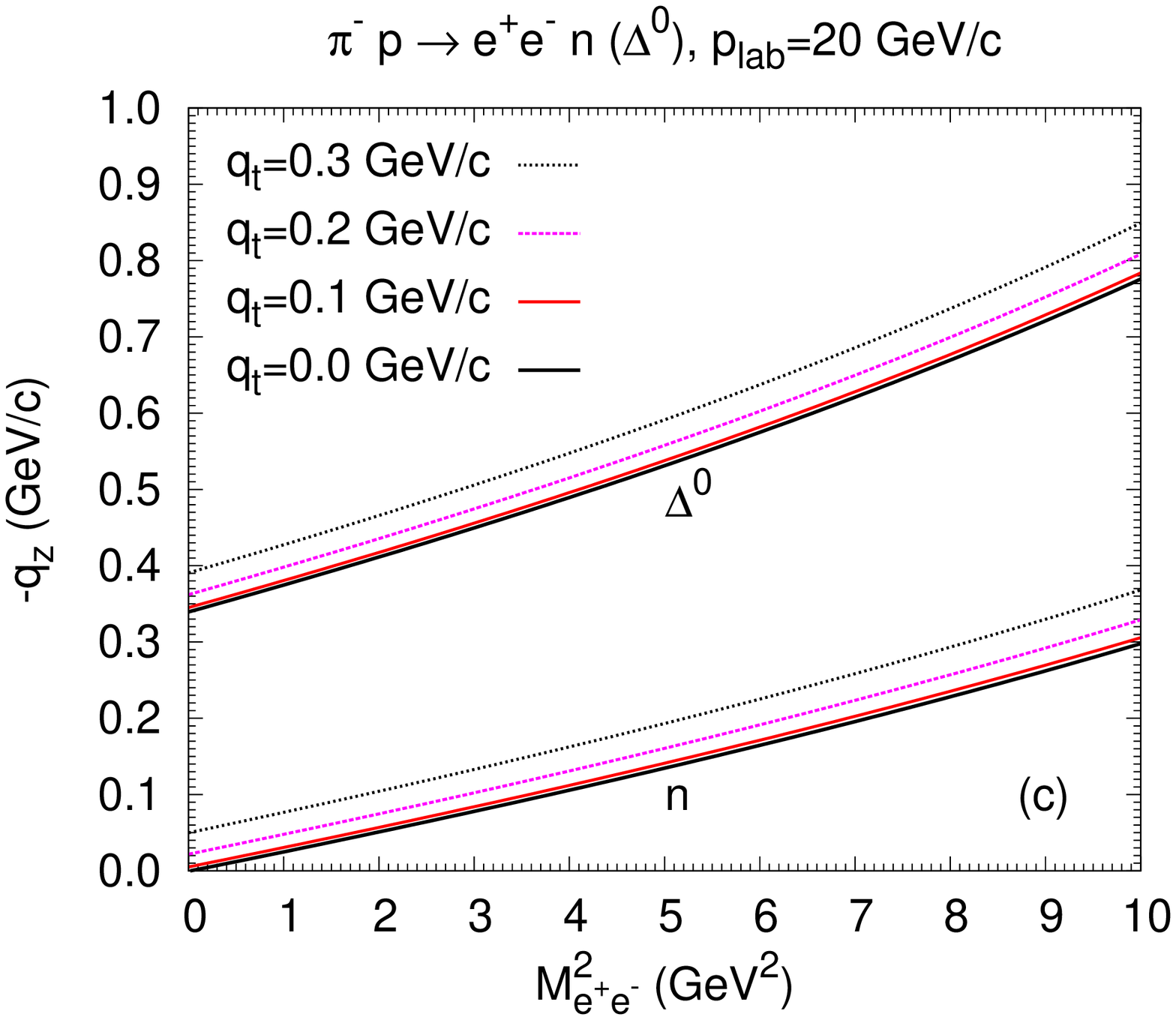}&
\\
\end{tabular}
\caption{\label{fig:kz_vs_m2ee} (color online) The longitudinal component of the momentum of outgoing neutron 
or $\Delta^0$ in the reaction $\pi^- p \to l^+l^- n(\Delta^0)$ on proton at rest as a function 
of the invariant mass squared of the $l^+l^-$ pair for several values of the transverse momentum transfer, $q_t$,
as indicated. Panels (a), (b) and (c) correspond to the beam momenta 9, 16 and 20 GeV/c.}
\end{figure}
The formula  (\ref{T_e+e-}) for the nuclear transparency is generally valid for the production mechanisms of the $l^+l^-$ pair, for example, 
with the $\Delta^0$ excitation, $\pi^- p \to l^+l^- \Delta^0$.
However, this channel probes the $p \to \Delta^0$ transition GPD's and not the $p \to n$ ones
and, thus, should be considered separately.
Moreover, it is also possible to produce a dilepton pair of a given invariant mass in a quite violent process $\pi^- p \to l^+l^- B$
with a large momentum transfer to any charge neutral nucleon or $\Delta$ resonance state $B$.

To understand the relevant kinematics better, Fig.~\ref{fig:kz_vs_m2ee} shows the longitudinal 
component of the momentum transfer to the target, $-q^z$, as a function of the invariant mass squared of the dilepton 
pair for several beam momenta. This is expressed by the relation
\begin{equation}
   M^2_{l^+l^-}=m^2_{\pi} + m_B^2 - m_p^2 - 2(E_{\pi}+m_p)(\sqrt{m_B^2+\mathbf{q}^2}-m_p) - 2 p_{\rm lab} q^z~,    \label{M^2_ee} 
\end{equation}
where $B=n(\Delta^0)$ for the outgoing neutron ($\Delta^0$). The reaction channel with the outgoing neutron can be
clearly separated with a pretty modest momentum resolution of the detector 
if one restricts the value of the longitudinal momentum transfer for the fixed value of $M^2_{l^+l^-}$. 
With increasing beam momentum the longitudinal momentum transfer for the production of dileptons with fixed invariant mass 
is decreasing. For example, for $M^2_{l^+l^-} = 4$ GeV$^2$, we see that the lowest longitudinal momentum transfer 
is 0.262, 0.135 and 0.107 GeV/c for $p_{\rm lab}=9, 16$ and $20$ GeV/c, respectively. Thus, at large beam momenta, 
$p_{\rm lab} \simeq 20$ GeV/c, the reaction should, indeed, be dominated by the formation of the excited states of the nucleus $A_{Z-1}$.

\section{Nuclear transparency}
\label{Transparency}

\subsection{Pion electroproduction at JLab}
\label{elePi}

\begin{figure}
\begin{tabular}{cc}
\includegraphics[scale = 0.45]{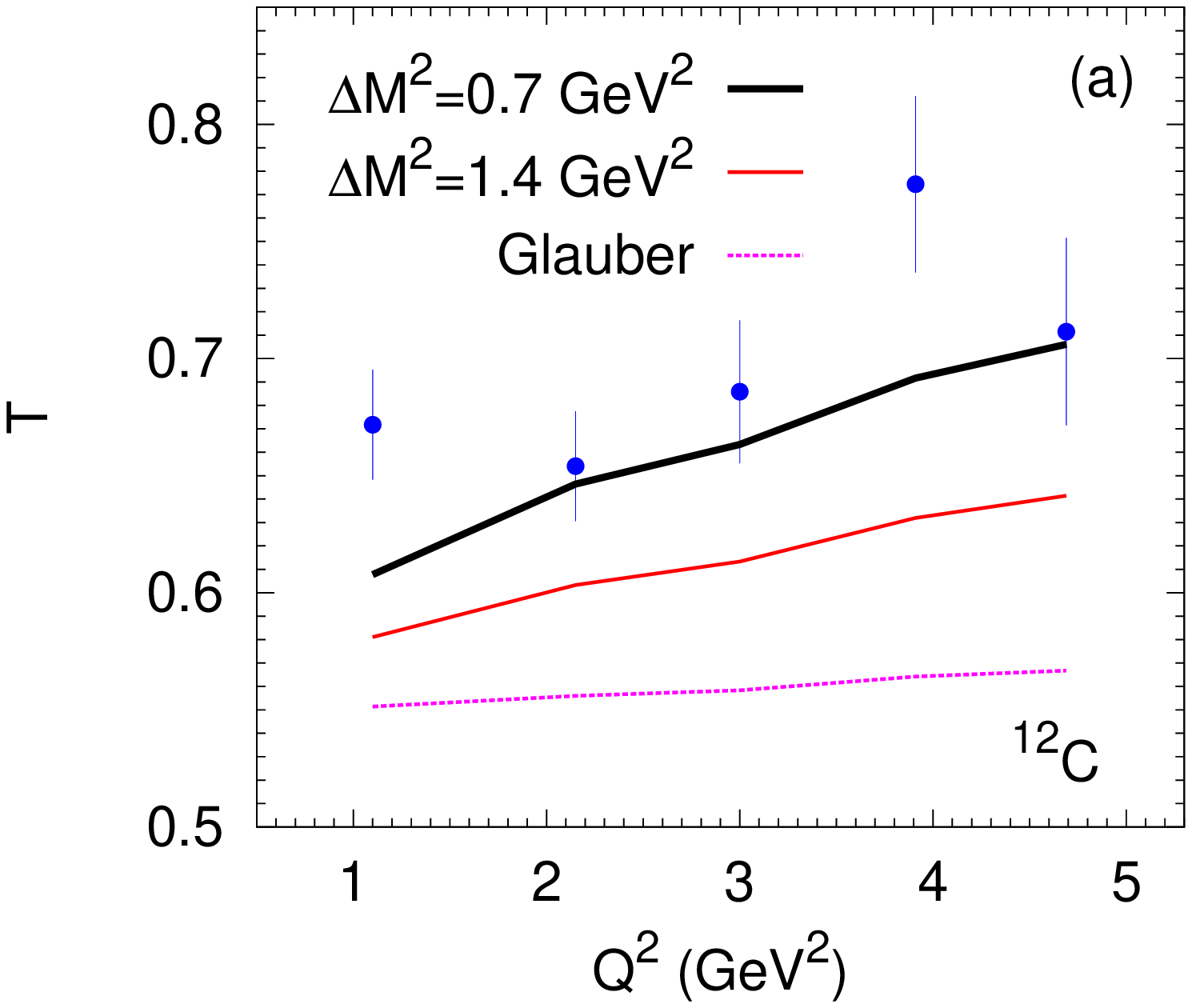} &
\includegraphics[scale = 0.45]{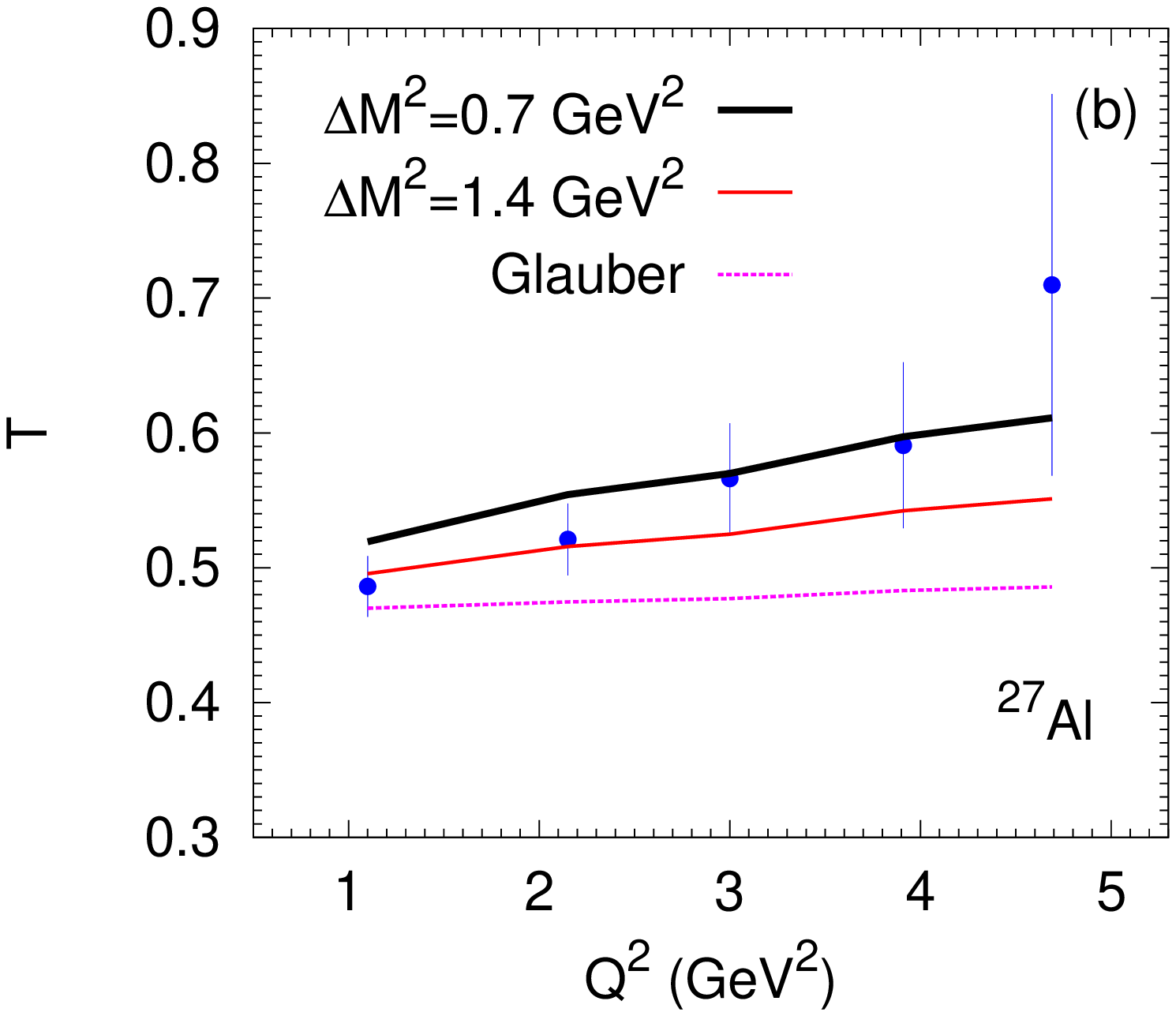} \\
\includegraphics[scale = 0.45]{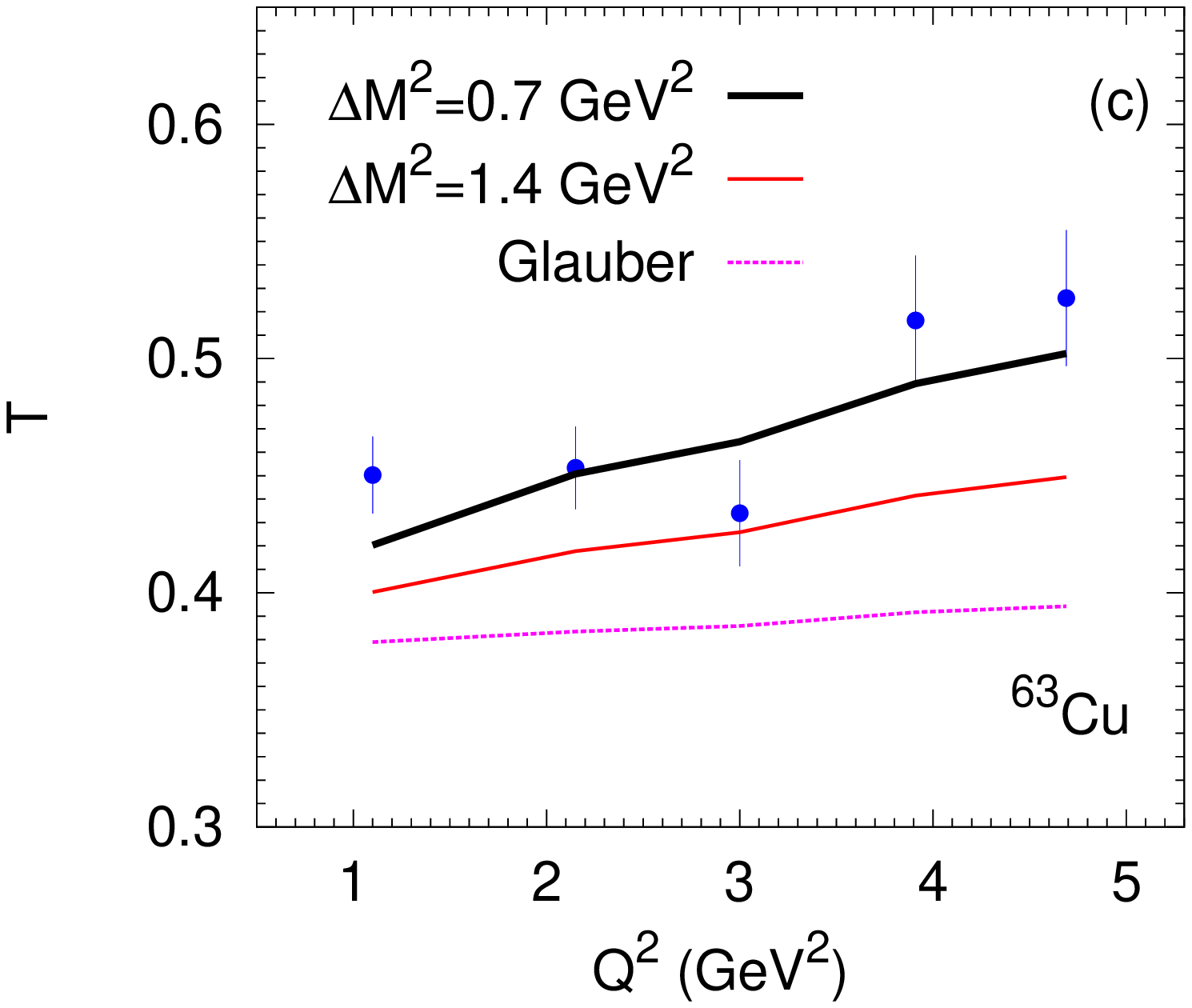} &
\includegraphics[scale = 0.45]{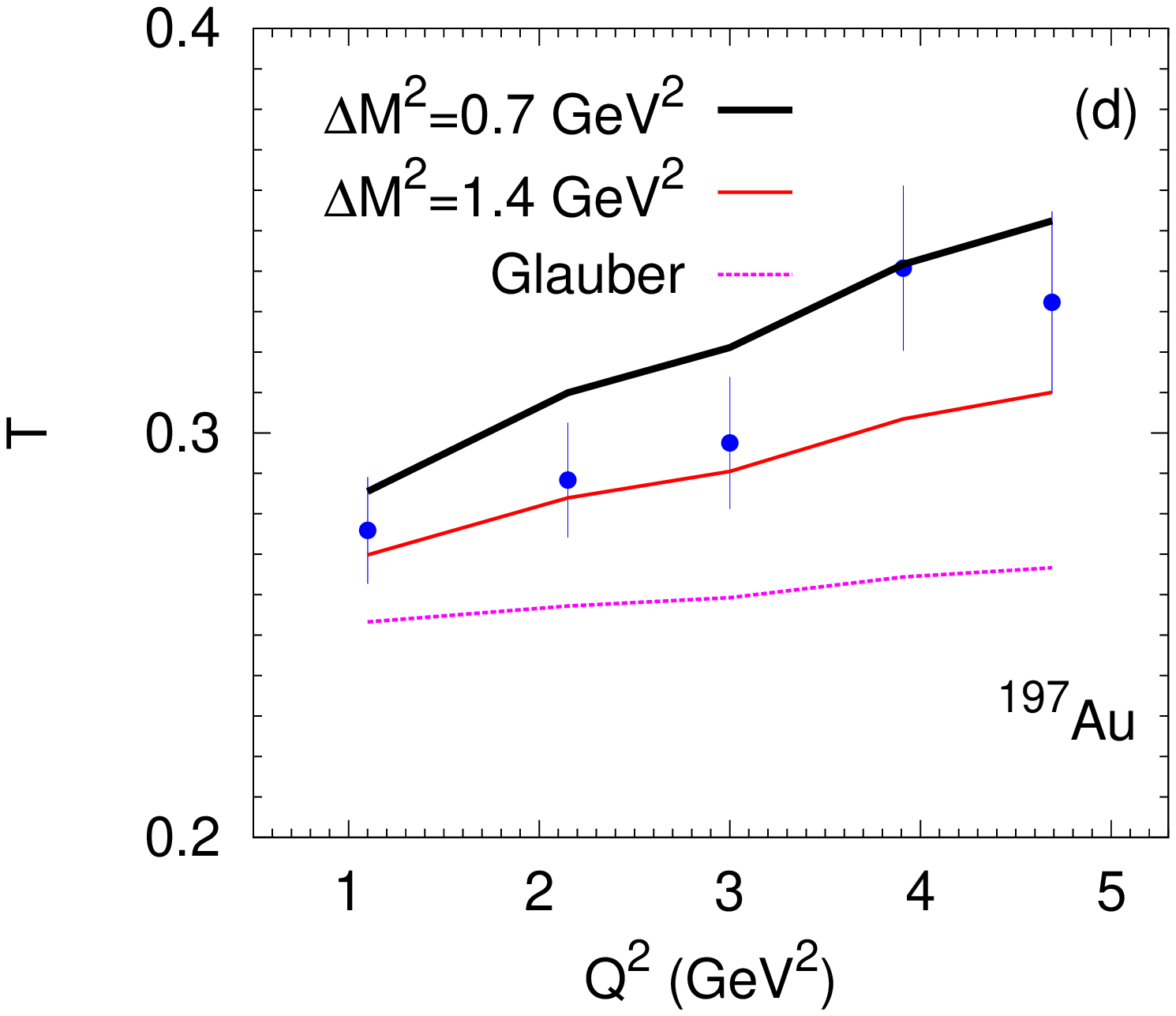} \\
\end{tabular}
\caption{\label{fig:T_JLab} (color online) Transparency
vs $Q^2$ for the $(e,e^\prime \pi^+)$ reaction on $^{12}$C, $^{27}$Al, $^{63}$Cu, and $^{197}$Au
targets (panels (a), (b), (c), and (d), respectively) in the collinear kinematics. Dashed lines -- Glauber model,
thick (thin) solid lines -- quantum diffusion model with $\Delta M^2=0.7 (1.4)$ GeV$^2$.
The values of pion momentum are chosen as $p_{\pi}=2.793, 3.187, 3.418, 4.077$, and $4.412$ GeV/c for 
$Q^2=1.10, 2.15, 3.00, 3.91$ and $4.69$, respectively, according to the experimental conditions \cite{Clasie:2007aa}.
The experimental data are from Ref. \cite{Clasie:2007aa}.}
\end{figure}
Before discussing predictions for the pion-induced reactions it is instructive to compare our present calculations
with available experimental data from JLab \cite{Clasie:2007aa} on nuclear transparency in the pion electroproduction 
reaction $A(e,e^\prime \pi^+)$.
JLab results \cite{Clasie:2007aa} were also described in the CT framework in Refs. \cite{Larson:2006ge,Cosyn:2007er}  
while an alternative description of the data using GiBUU model was reported in Ref. \cite{Kaskulov:2008ej}.

In the kinematics of the JLab experiment \cite{Clasie:2007aa}, the outgoing pion momentum is almost parallel to the momentum of the virtual
photon and the momentum transfer to the nucleus is smaller than 1 GeV/c.
Thus, the kinematics of this experiment selects the $p \to n$ transition in the nuclear target. 
The corresponding formula for the nuclear transparency is given by Eq.(\ref{T_e+e-}) with the only change that the integration 
over the outgoing pion trajectory from $z$ to $+\infty$ has to be done in the pion survival probability (cf. Ref. \cite{Larson:2006ge})
\footnote{Due to the parity conservation by strong interaction, the nucleon density distributions are always invariant under reflection
about the nuclear centre,  $\rho_q(-\mathbf{r})=\rho_q(\mathbf{r})$, $q=n,p$. Hence, the nuclear transparency for the
pion electroproduction is identically equal to the nuclear transparency of Eq.(\ref{T_e+e-}) for similar momenta of incoming and outgoing pions
(if the sizes of the $q\bar q$ in the interaction point are the same).}.
The transparency for the $(e,e^\prime \pi^+)$ reaction is shown in Fig.~\ref{fig:T_JLab} as a function of the momentum transfer squared.
One clearly observes that the calculations within the quantum diffusion model (see Eqs.(\ref{sigma_piN_eff}),(\ref{Psurv^CT})) are in better 
agreement with the data than the Glauber model calculation. The Glauber calculation produces a too flat $Q^2$-dependence of the transparency 
and underestimates the experimental data.
The default pion coherence length Eq.(\ref{l_pi}) calculated with $\Delta M^2 = 0.7$ GeV$^2$
describes well the transparency for all studied nuclei except the heaviest one, $^{197}$Au, where the calculation with 
a two times shorter coherence length seems to agree better with the data.
The pion momenta in the JLab experiment \cite{Clasie:2007aa} are between 2.8 and 4.4 GeV/c which corresponds 
to a pion coherence length between 1.6 and 2.5 fm. This is comparable with the r.m.s. charge radii of light nuclei, 
$\langle r^2 \rangle^{1/2}_{^{12}\rm C}=2.46$ fm, $\langle r^2 \rangle^{1/2}_{^{27}\rm Al}=3.05$ fm, however, significantly less than the radii 
of the heavy ones, $\langle r^2 \rangle^{1/2}_{^{63}\rm Cu}=3.93$ fm, $\langle r^2 \rangle^{1/2}_{^{197}\rm Au}=5.33$ fm \cite{DeJager:1974dg}. 
The relative effect of the CT on the pion survival probability 
at fixed hard interaction point $(\mathbf{b},z)$ can be expressed as 
\begin{equation}
    \frac{{\cal P}_{\pi,{\rm surv}}^{\rm CT}(\mathbf{b},z,+\infty)}{{\cal P}_{\pi,{\rm surv}}(\mathbf{b},z,+\infty)} 
    = \exp\left(\int\limits_z^{z+l_{\pi}}dz^\prime (\sigma_{\pi N}(p_{\pi}) -\sigma_{\pi N}^{\rm eff}(p_{\pi},z^\prime - z)) \rho(\mathbf{b},z^\prime)\right)~.
                                              \label{P_surv_ratio}
\end{equation}
Since for heavier nuclei the integral over the pion trajectory picks up on-average larger values of the nuclear density, the relative effect of CT 
is stronger for heavier nuclei as also noted in Ref. \cite{Larson:2006ge}.

\subsection{Dilepton production in $\pi^-$-nucleus reactions}
\label{piDil}

\begin{figure}
\begin{tabular}{cc}
\includegraphics[scale = 0.5]{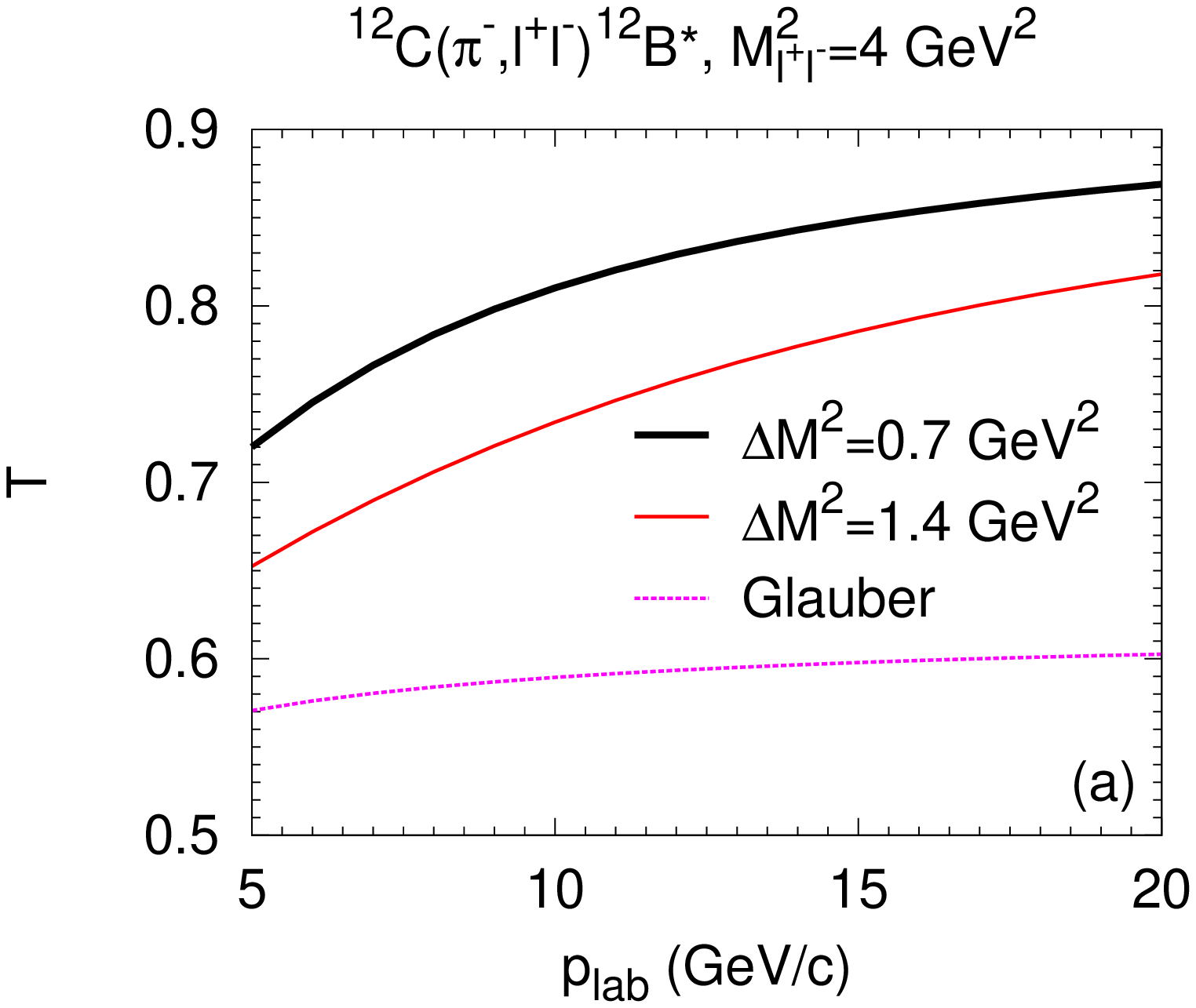} &
\includegraphics[scale = 0.5]{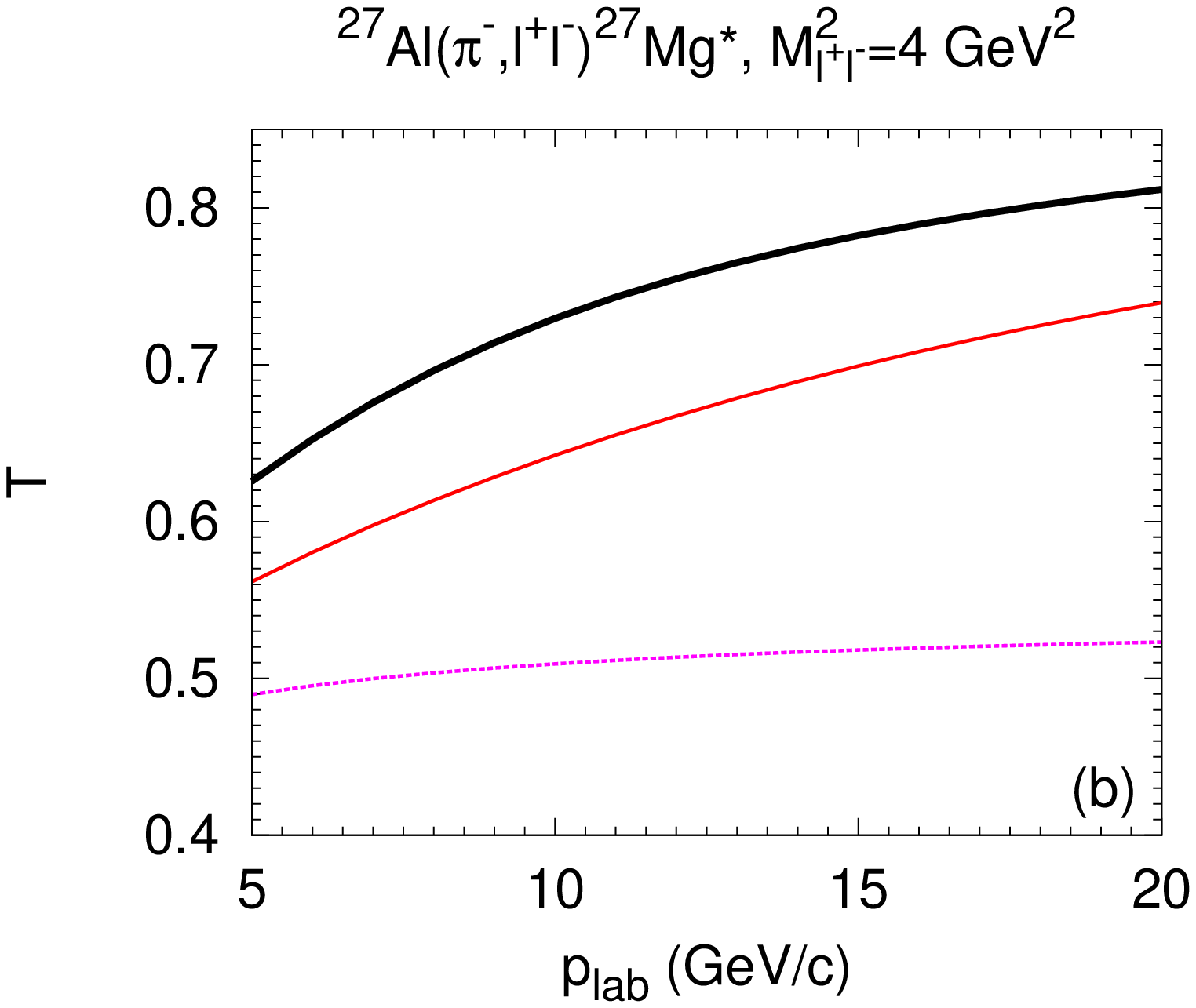} \\
\includegraphics[scale = 0.5]{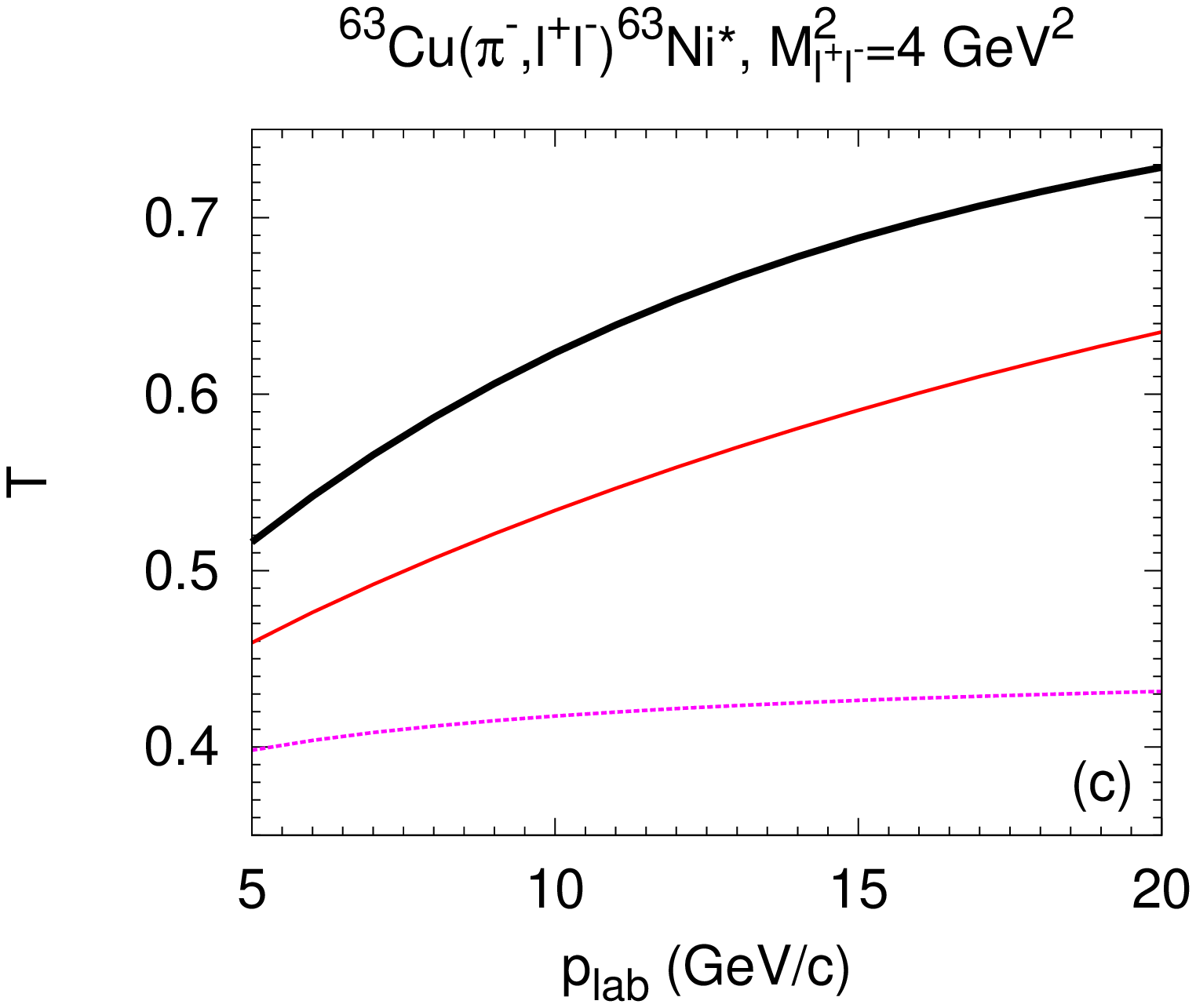} &
\includegraphics[scale = 0.5]{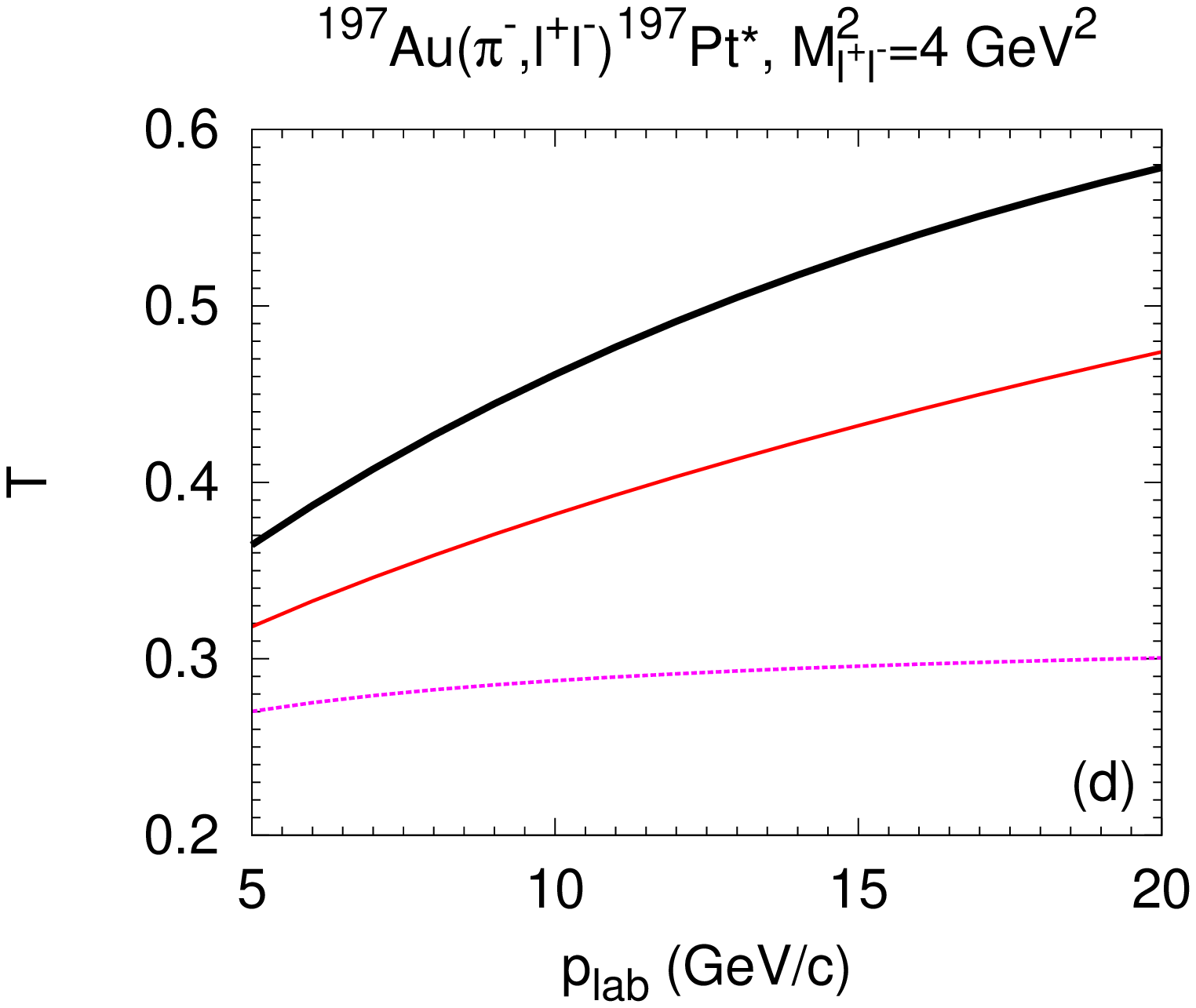} \\
\end{tabular}
\caption{\label{fig:T_4gev2} (color online) Transparency vs pion beam momentum for the $(\pi^-,l^+l^-)$ reaction 
at fixed $M^2_{l^+l^-}=4$ GeV$^2$ on $^{12}$C, $^{27}$Al, $^{63}$Cu, and $^{197}$Au targets 
(panels (a), (b), (c), and (d), respectively). 
Dashed lines -- Glauber model, thick (thin) solid lines -- quantum diffusion model with $\Delta M^2=0.7~(1.4)$ GeV$^2$.}
\end{figure}
Let us now discuss pion-induced reactions. Figure~\ref{fig:T_4gev2} shows the transparency 
for the $(\pi^-,l^+l^-)$ reaction as a function of the pion beam momentum at fixed invariant mass of the dilepton pair.
We observe strong effects of CT on the transparency that are even  
more pronounced for heavier nuclei, similar to the case of pion electroproduction. 
The beam momentum dependence at fixed resolution scale, $M^2_{l^+l^-}$,
allows, however, to get more information on the pion coherence length. As we observe, in calculations within 
the quantum diffusion model the transparency varies by $\sim 5\%$ for $^{12}$C and by $\sim 20\%$ for $^{197}$Au in the beam momentum range 15-20 GeV/c 
and it is much larger than in the Glauber model.
Thus, the transparency 
saturates at smaller $p_{\rm lab}$ for lighter nuclei than for the heavier ones.
This is because the saturation is reached when the pion coherence length becomes comparable with the nuclear diameter. 

\begin{figure}
\begin{tabular}{cc}
\includegraphics[scale = 0.5]{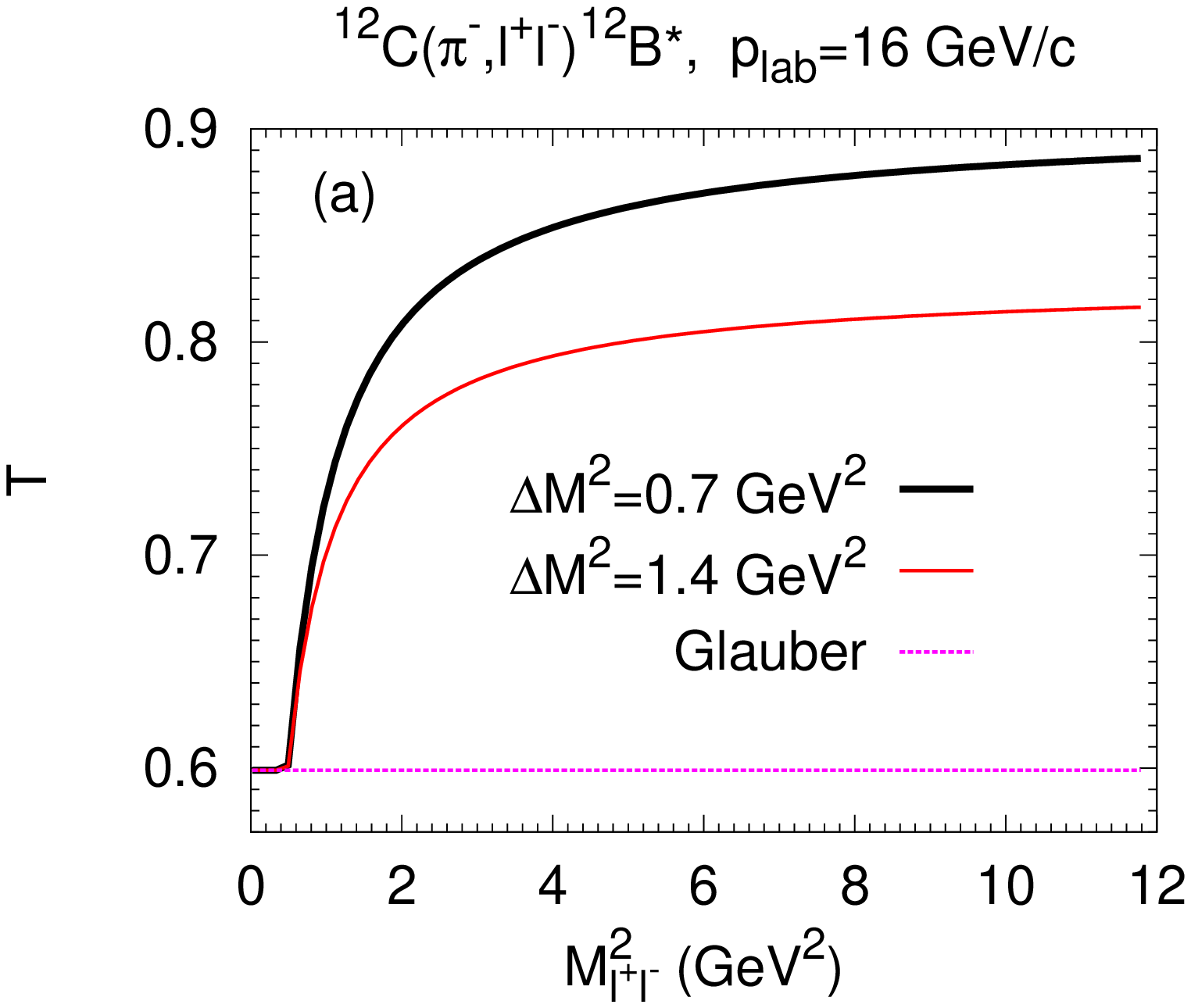} &
\includegraphics[scale = 0.5]{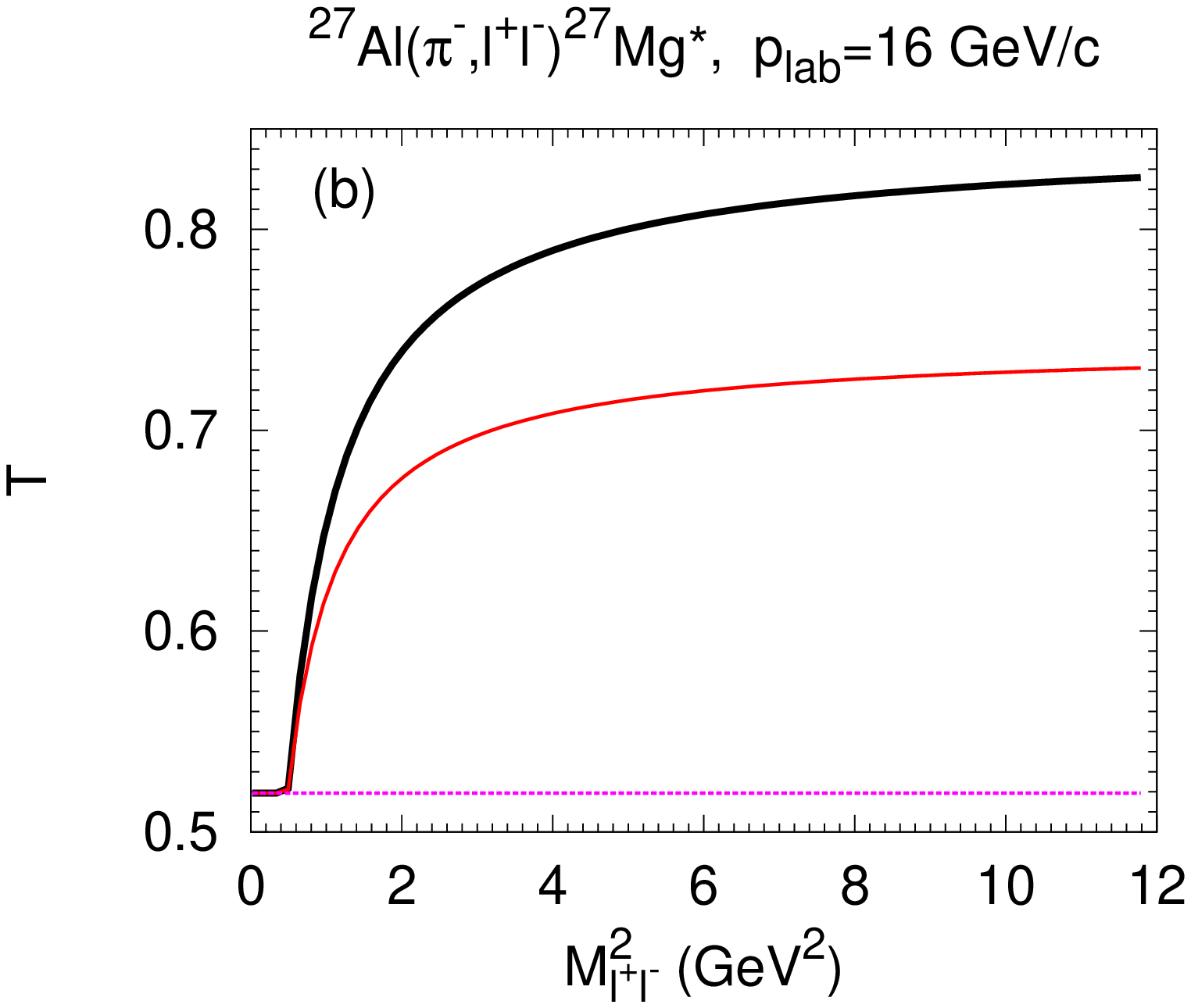} \\
\includegraphics[scale = 0.5]{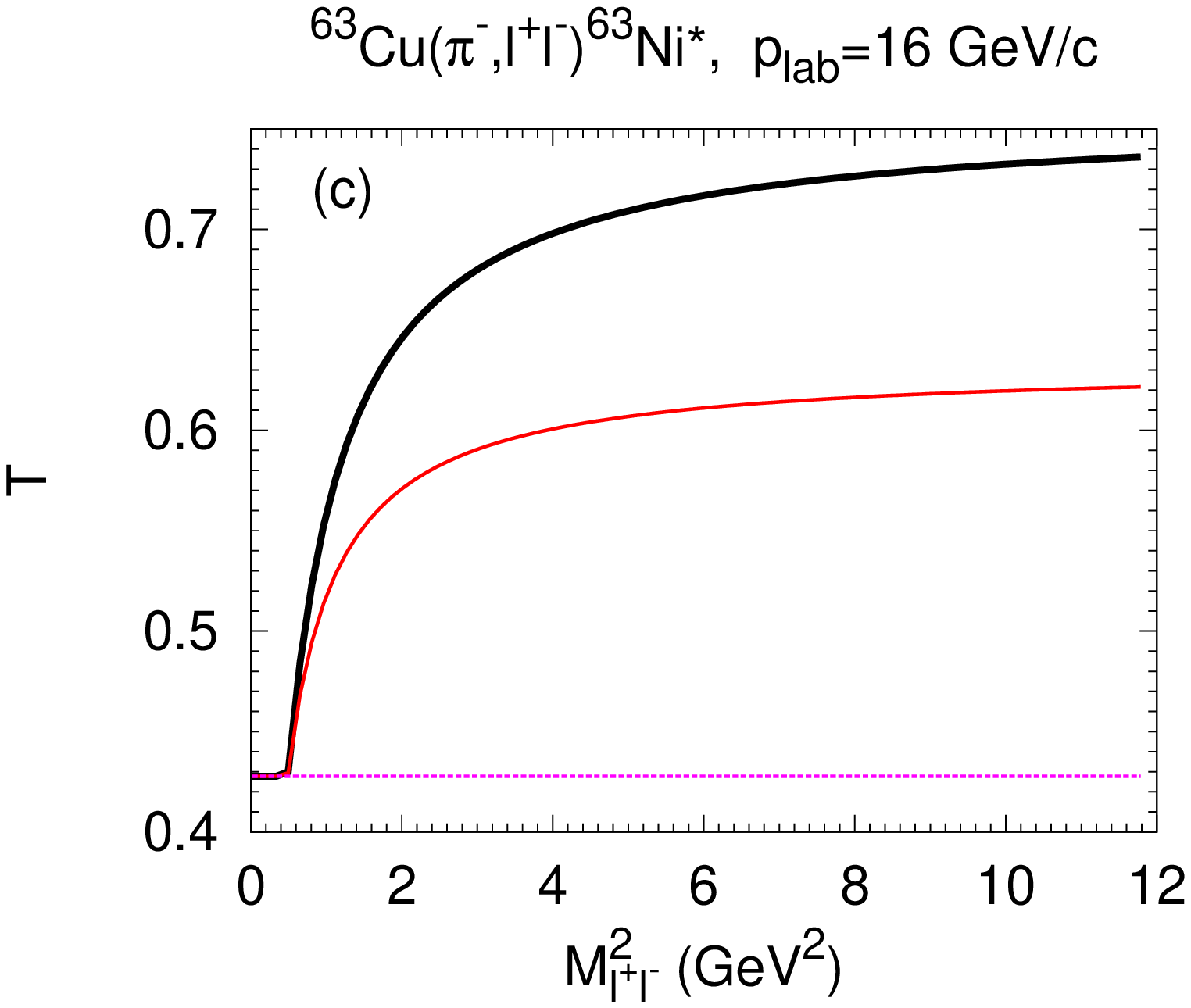} &
\includegraphics[scale = 0.5]{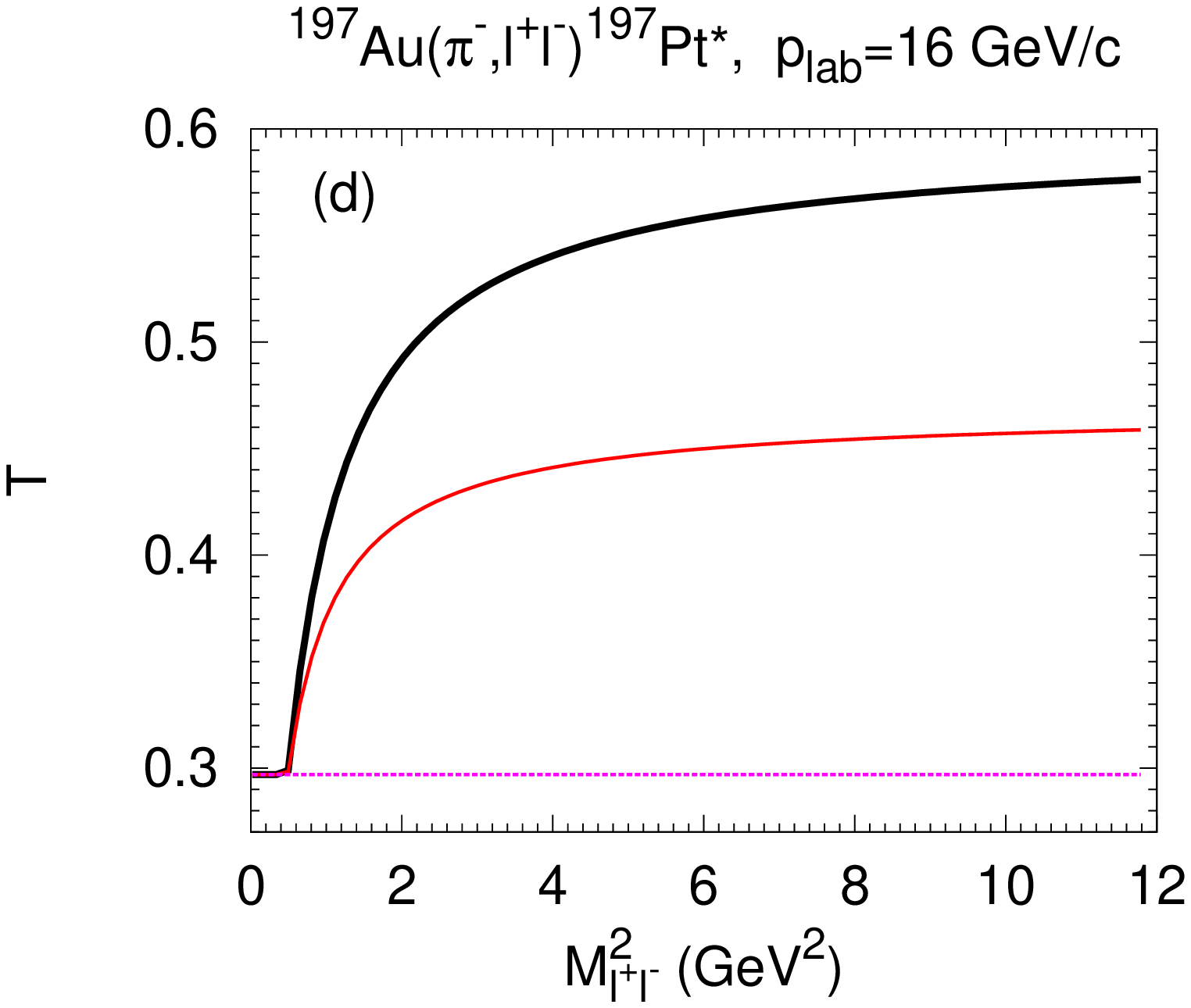} \\
\end{tabular}
\caption{\label{fig:T_16gevc} (color online) Transparency 
vs invariant mass squared of the $l^+l^-$ pair for the $(\pi^-,l^+l^-)$ reaction 
at $p_{\rm lab}=16$ GeV/c on $^{12}$C, $^{27}$Al, $^{63}$Cu, and $^{197}$Au targets
(panels (a), (b), (c), and (d), respectively). The lines have the same meaning
as in Fig.~\ref{fig:T_4gev2}.}
\end{figure}
In Fig.~\ref{fig:T_16gevc} we show the transparency 
as a function of the square of the invariant mass of the dilepton pair at fixed beam momentum. 
Independent on the choice of the target nucleus, the transparency 
saturates at $M^2_{l^+l^-} \simeq 4$ GeV$^2$. The sensitivity to the pion coherence length is stronger for heavier
nuclei, since $l_{\pi}=9~(4.5)$ fm for $\Delta M^2=0.7~(1.4)$ GeV$^2$ at 16 GeV/c. Thus, $l_{\pi}$ is comparable with or larger 
than the $^{12}$C diameter, however, $l_{\pi}$ is always less than the $^{197}$Au diameter. In other words, the pion interacting 
with the $^{12}$C nucleus is almost in the point-like configuration across its entire trajectory within the nucleus, 
while for the $^{197}$Au nucleus the pion dynamics is still in the region of quantum diffusion.

\begin{figure}
\includegraphics[scale = 0.6]{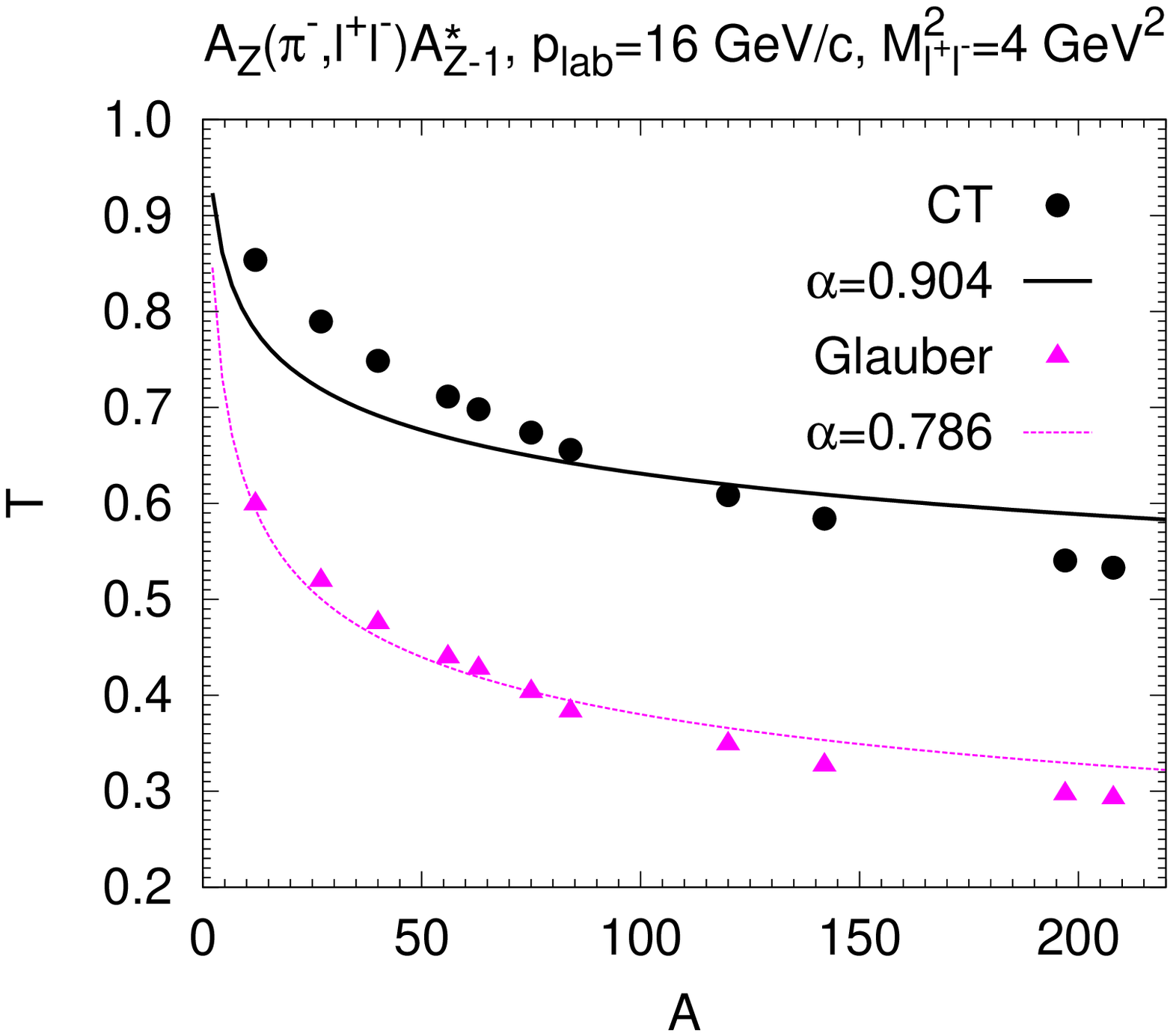}
\caption{\label{fig:T_16gevc_4gev2} (color online) Target nucleus mass number dependence of the transparency 
for the $A_Z(\pi^-,l^+l^-)A^*_{Z-1}$ reaction at $p_{\rm lab}=16$ GeV/c and $M^2_{l^+l^-}=4$ GeV$^2$.
The calculations are done for nuclei $^{12}$C, $^{27}$Al, $^{40}$Ca, $^{56}$Fe, $^{63}$Cu, $^{75}$As, $^{84}$Kr, 
$^{120}$Sn, $^{142}$Ce, $^{197}$Au, and $^{208}$Pb. The result obtained within the quantum diffusion model 
is shown by solid circles while the Glauber model result is shown by solid triangles. 
The fits with a power law of Eq.(\ref{T_vs_A}) are shown by lines marked with the value of $\alpha$.}
\end{figure}
Fig.~\ref{fig:T_16gevc_4gev2} shows the mass dependence of the transparency 
for the $(\pi^-,l^-l^+)$ reaction at fixed values of 
the beam momentum, $p_{\rm lab}=16$ GeV/c, and $l^+l^-$ invariant mass, $M^2_{l^+l^-}=4$ GeV$^2$. 
We performed the fits of the calculated mass number dependence of the transparency by the power law
\begin{equation}
    T \simeq A^{\alpha-1}~.                             \label{T_vs_A}
\end{equation}
The Glauber model results can be well reproduced with $\alpha=0.786 \pm 0.003$. 
This is not so far from the surface dominated production mechanism ($\alpha=2/3$).  
The power law fit of the quantum diffusion calculation is less good, although it is clear that the obtained value,
$\alpha=0.904 \pm 0.005$, is rather close to the volume dominated production ($\alpha=1$).

\begin{figure}
\includegraphics[scale = 0.6]{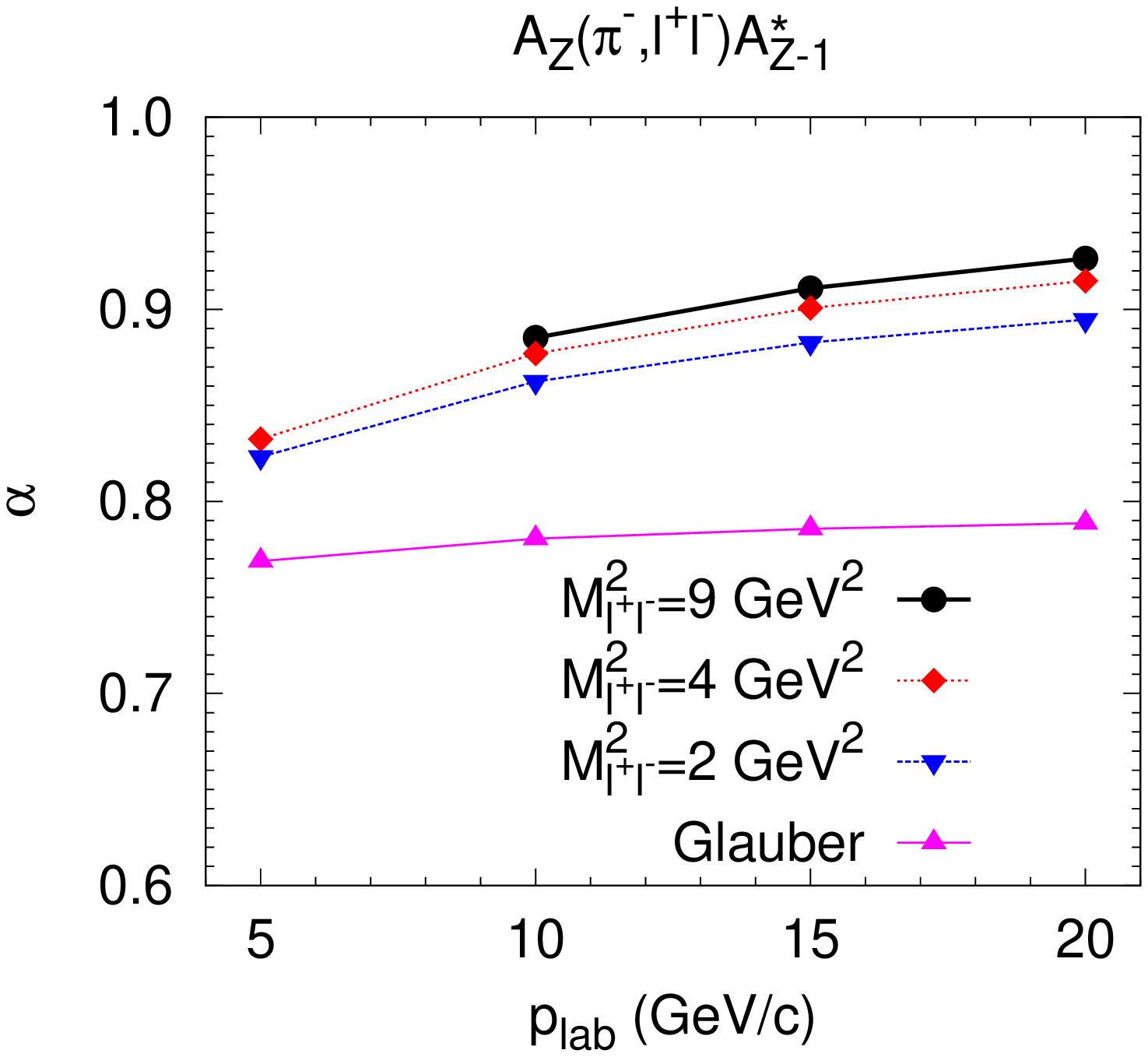}
\caption{\label{fig:alpha} (color online) Beam momentum and dilepton invariant mass dependence of the parameter $\alpha$ in
the power law fit (\ref{T_vs_A}) of the transparency for the $A_Z(\pi^-,l^+l^-)A^*_{Z-1}$ reaction.
The quantum diffusion model calculations for the different values of $M^2_{l^+l^-}$ as indicated and
the Glauber model results are shown.}
\end{figure}
In Fig.~\ref{fig:alpha} we display the parameter $\alpha$ of the power law fit (\ref{T_vs_A}) as a function of the beam momentum
for several values of $M^2_{l^+l^-}$. At the lowest $p_{\rm lab}=5$ GeV/c the results of the quantum diffusion model calculations 
and the Glauber model results are close to those of Fig. 3 of Ref. \cite{Clasie:2007aa} at $Q^2=4$ GeV$^2$ which can be regarded as a benchmark.
There is a significant, $\sim 15\%$, enhancement of the slope parameter $\alpha$ at the largest $p_{\rm lab}=20$ GeV/c due to the CT.
The enhancement depends only weakly on the dilepton invariant mass at $M^2_{l^+l^-} \gtsim 4$ GeV$^2$ (see also Fig.~\ref{fig:T_16gevc}).

\subsection{Vector meson production in $\pi^-$-nucleus reactions}
\label{piVec}

\begin{figure}
\begin{tabular}{cc}
\includegraphics[scale = 0.5]{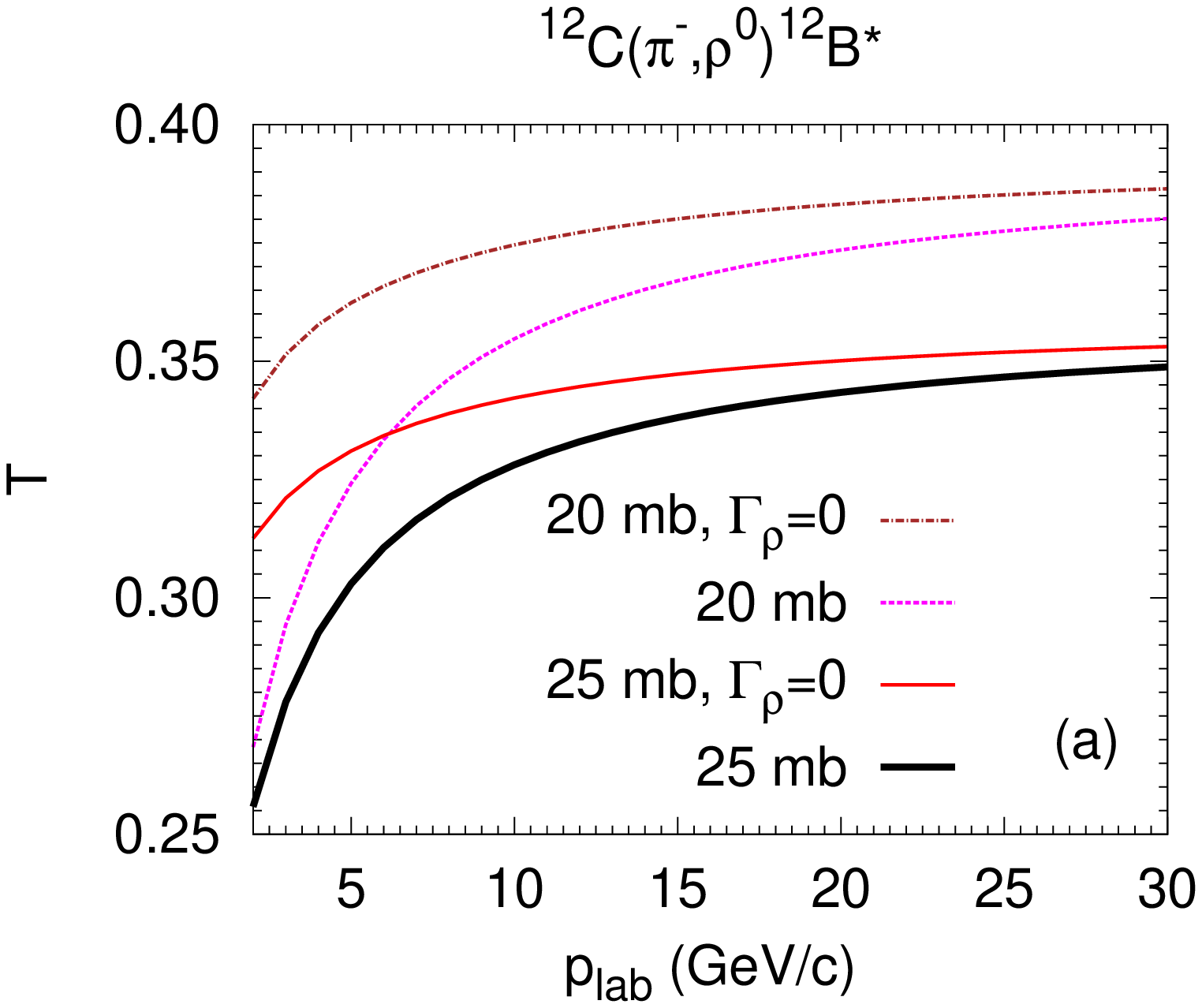} &
\includegraphics[scale = 0.5]{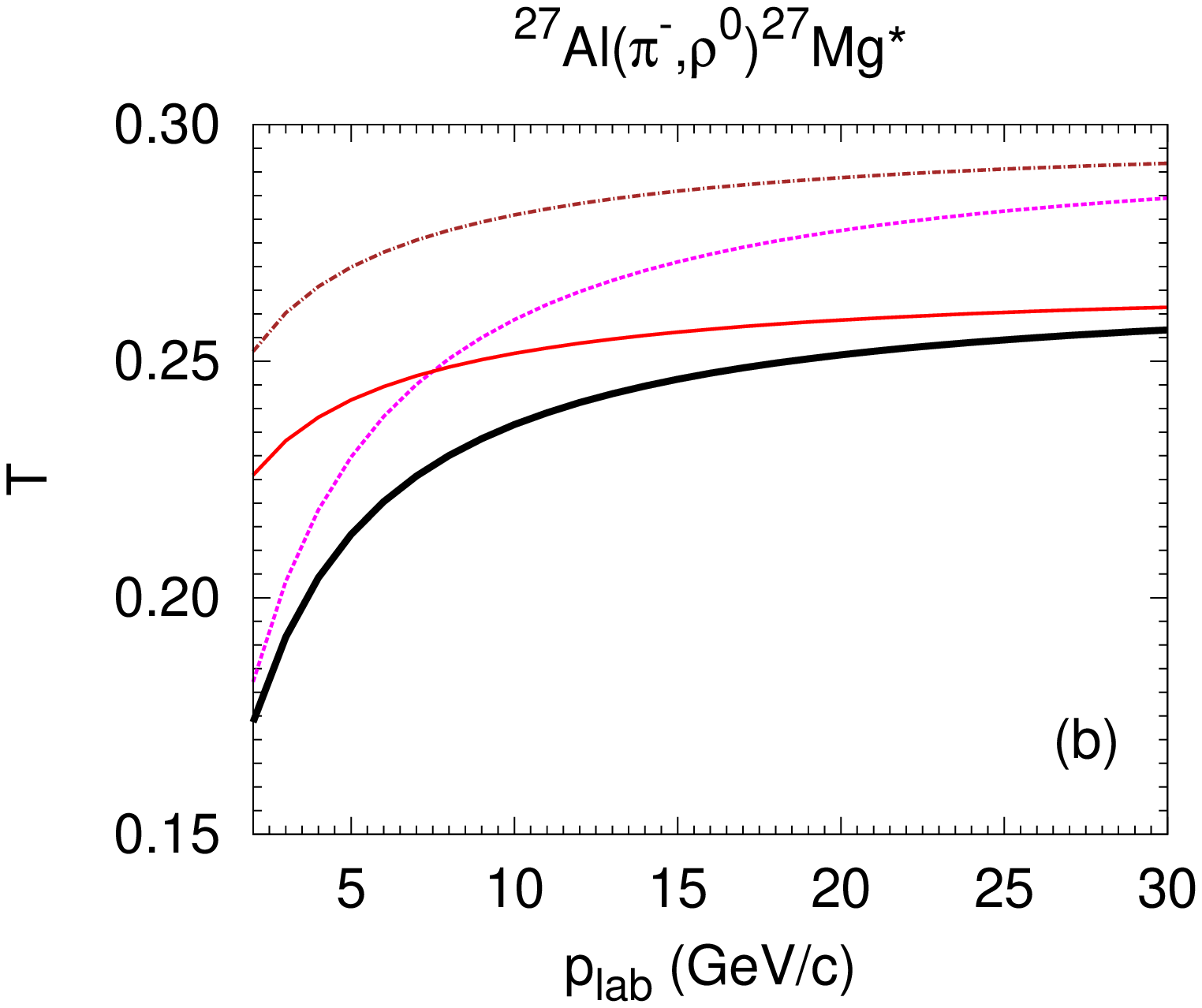} \\
\includegraphics[scale = 0.5]{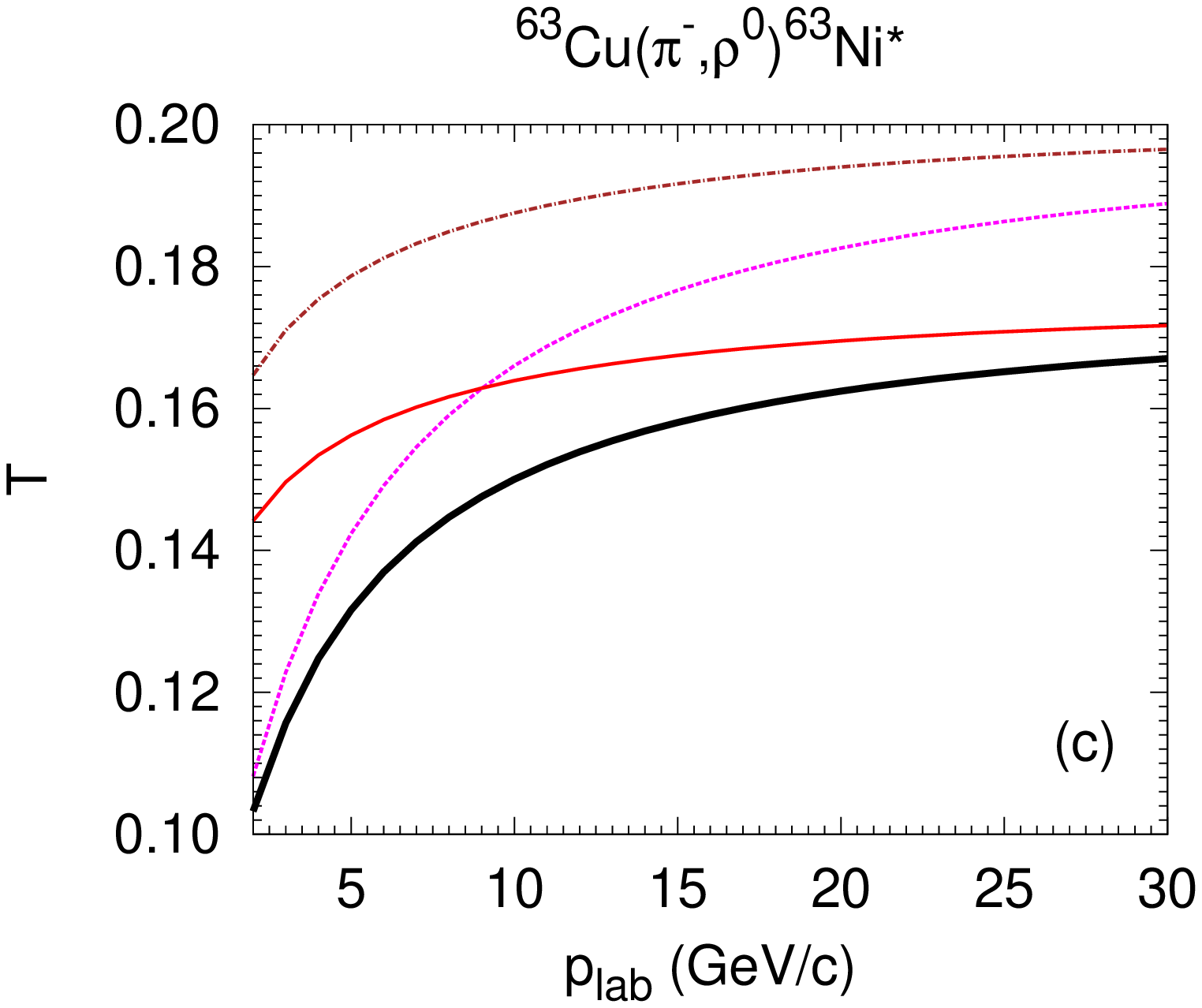} &
\includegraphics[scale = 0.5]{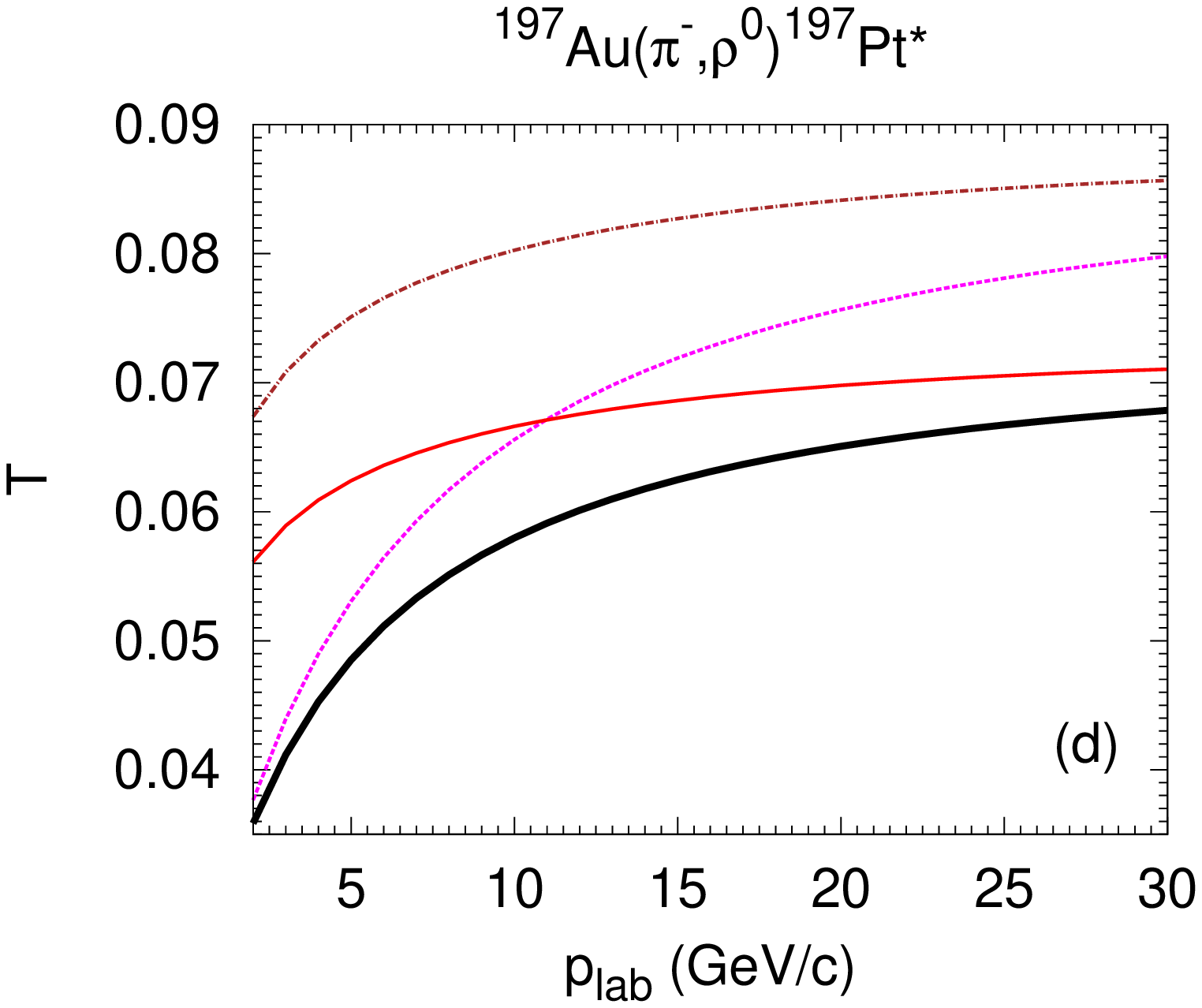} \\
\end{tabular}
\caption{\label{fig:T_pi_rho} (color online) Transparency vs beam momentum for the $(\pi^-,\rho^0)$ reaction 
on $^{12}$C, $^{27}$Al, $^{63}$Cu, and $^{197}$Au targets (panels (a), (b), (c), and (d), respectively).
The calculations are performed with and without  
$\rho$ decay inside nucleus by setting the total $\rho N$ cross section equal to 25 and 20 mb
as indicated.}
\end{figure}
The transparency 
in $\pi^-$-induced $\rho^0$ production reactions on different nuclei is shown in
Fig.~\ref{fig:T_pi_rho} as a function of beam momentum. Without taking into account $\rho$-decays inside the
nucleus (i.e. setting $\Gamma_\rho=0$), the transparency is practically constant within the beam momentum range 10-30 GeV/c.
A small increase of the transparency $T$ is caused by slowly decreasing pion-nucleon cross section with increasing beam momentum, 
reaching a minimum of $\simeq 23$ mb at $p_{\rm lab} \simeq 50$ GeV/c. 
The $\rho$-decay inside the nucleus effectively leads to the replacement of the $\rho^0$ by a $\pi^+ \pi^-$ pair
which has a total cross section approximately two times larger than the $\rho N$ cross section (Eq.(\ref{sigma_rhoN})). 
This strongly reduces the transparency at low beam momenta. The reduction
is especially strong for the heavy nucleus, $^{197}$Au, reaching more than $20\%$ at $p_{\rm lab} \ltsim 10$ GeV/c.
With increasing beam momentum the influence of the $\rho$-decay becomes weaker due to the Lorentz factor
in Eq.(\ref{sigma_rhoN}). Thus, at $p_{\rm lab} \gtsim 15$ GeV/c the transparency is mostly defined by the value
of the total $\rho N$ cross section. 
To illustrate the sensitivity to $\sigma_{\rho N}$ we display in Fig.~\ref{fig:T_pi_rho} the results obtained with $\sigma_{\rho N}=20$ and
25 mb. We see that at the highest beam momentum the effect of $\rho N$ cross section variation on the transparency is $\sim 10\%$ for $^{12}$C
and $\sim 20\%$ for $^{197}$Au. However, the distortion of the transparency due to the $\rho$ decay is smaller for the light nuclei. 
Thus, the light nuclei seem to be better suited for the studies of the genuine $\rho$ interactions with nucleons at the beam momenta of few 10 GeV/c.

\begin{figure}
\includegraphics[scale = 0.6]{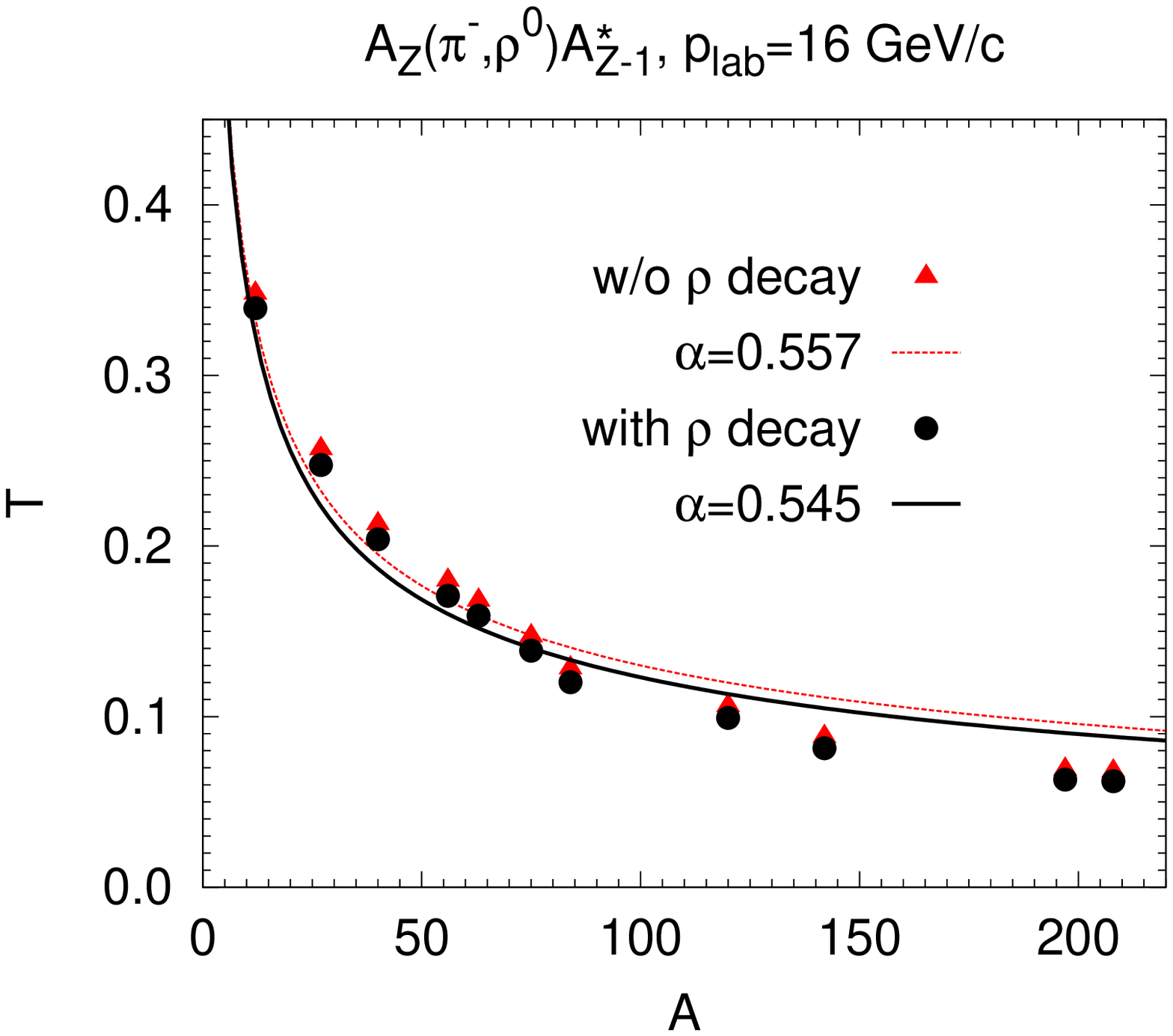}
\caption{\label{fig:T_16gevc_pi_rho} (color online) Target nucleus mass number dependence of the transparency  
for the $A_Z(\pi^-,\rho^0)A^*_{Z-1}$ reaction at $p_{\rm lab}=16$ GeV/c. 
The results with (without) $\rho$-decay inside nucleus are shown by solid circles (triangles).
The fits by a power law (\ref{T_vs_A}) are shown by lines labelled by the value of $\alpha$. 
The calculations are done for the same nuclei as in Fig.~\ref{fig:T_16gevc_4gev2}.} 
\end{figure}
The target nucleus mass dependence of the transparency for the $(\pi^-,\rho^0)$ process at $p_{\rm lab}=16$ GeV/c 
is shown in Fig.~\ref{fig:T_16gevc_pi_rho}. The difference between calculations with and without
$\rho$-decay is less than $10\%$ at this beam momentum (see also Fig.~\ref{fig:T_pi_rho}).
Both calculations can be reasonably well described by a power law, cf. Eq.(\ref{T_vs_A}), where
$\alpha=0.545 \pm 0.009$ (with $\rho$-decay) and $\alpha=0.557 \pm 0.009$  (without $\rho$-decay).
This indicates the surface character of the production mechanism.

\begin{figure}
\includegraphics[scale = 0.6]{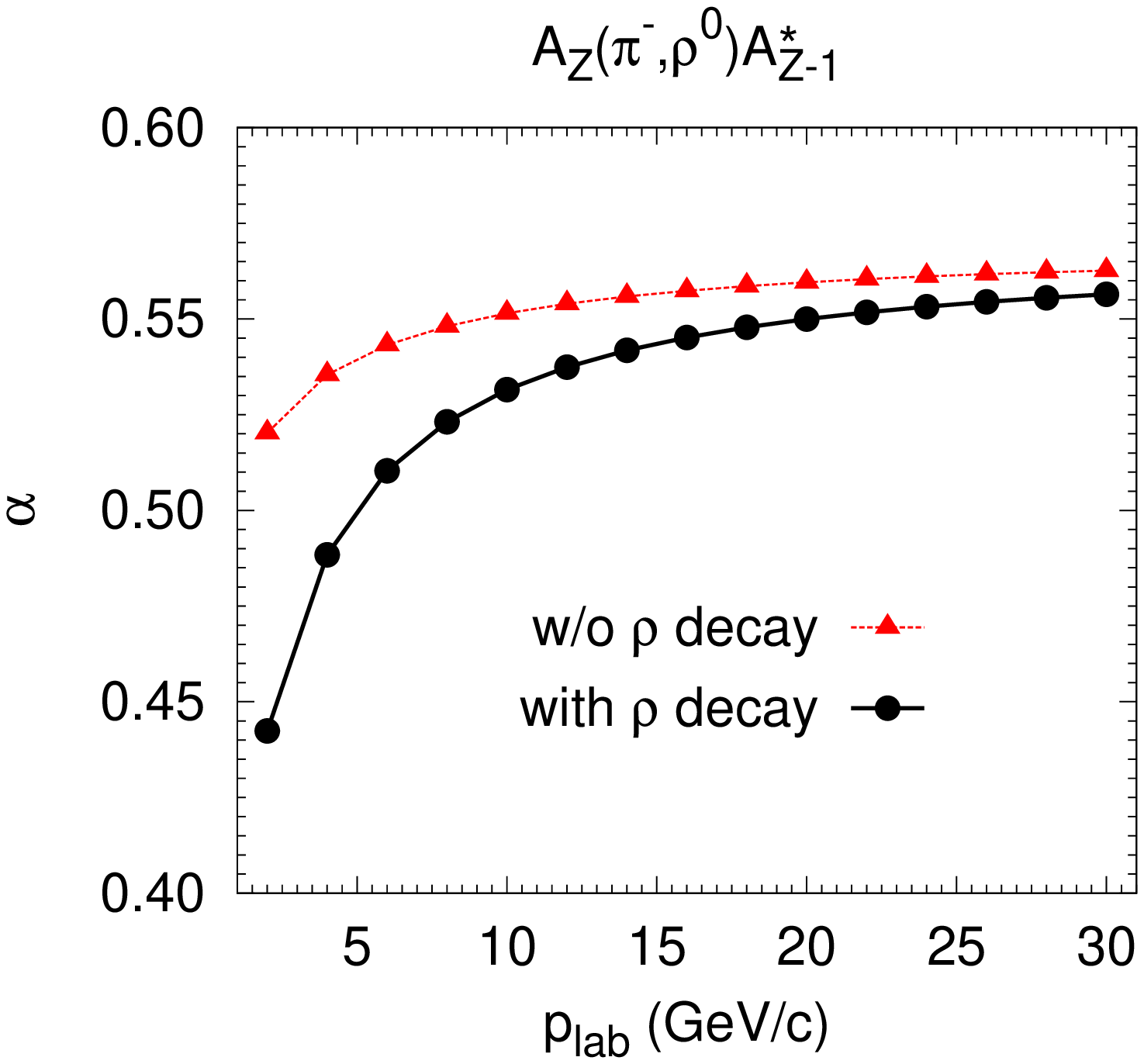}
\caption{\label{fig:alpha_pi_rho} (color online) Beam momentum dependence of the parameter $\alpha$ of the
power law fit (\ref{T_vs_A}) of the transparency for the $A_Z(\pi^-,\rho^0)A^*_{Z-1}$ reaction
calculated with (line with solid circles) and without (line with solid triangles) taking into account 
$\rho$ decay inside nucleus.}
\end{figure}
In Fig.~\ref{fig:alpha_pi_rho} we present the beam momentum dependence of the parameter $\alpha$ of the
power law fit for the $(\pi^-,\rho^0)$ process. In calculations without $\rho$-decay, $\alpha$ increases
by only 8\%, while the beam momentum grows from 2 to 30 GeV/c. Including $\rho$-decays inside the nucleus effectively
increases the absorption of $\rho$'s at low beam momenta, but only slightly alter the absorption of the high-momentum
$\rho$'s. This makes the parameter  $\alpha$ to increase much stronger, by up to 30\%, with the beam momentum.

\begin{figure}
\begin{tabular}{cc}
\includegraphics[scale = 0.4]{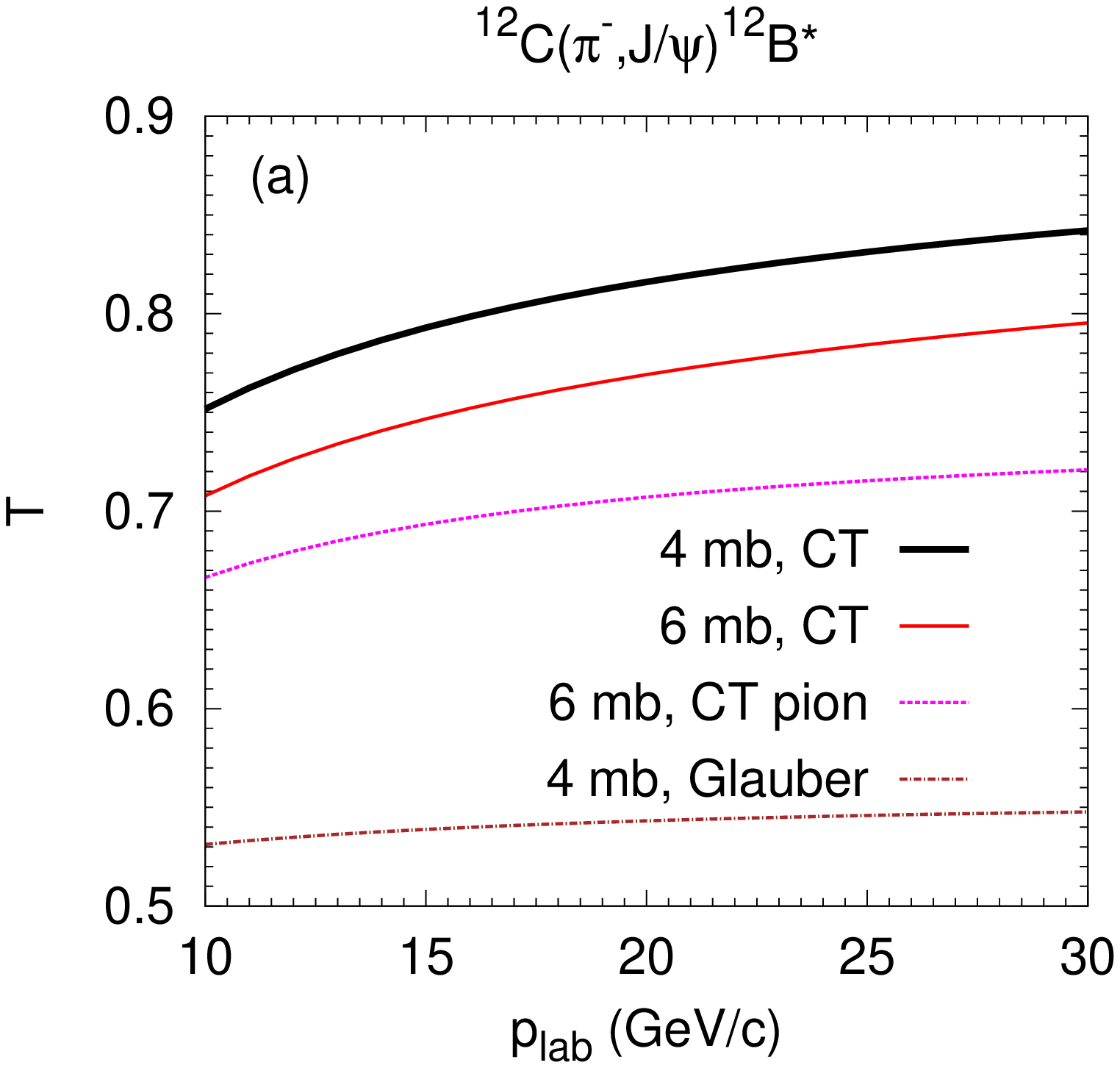} &
\includegraphics[scale = 0.4]{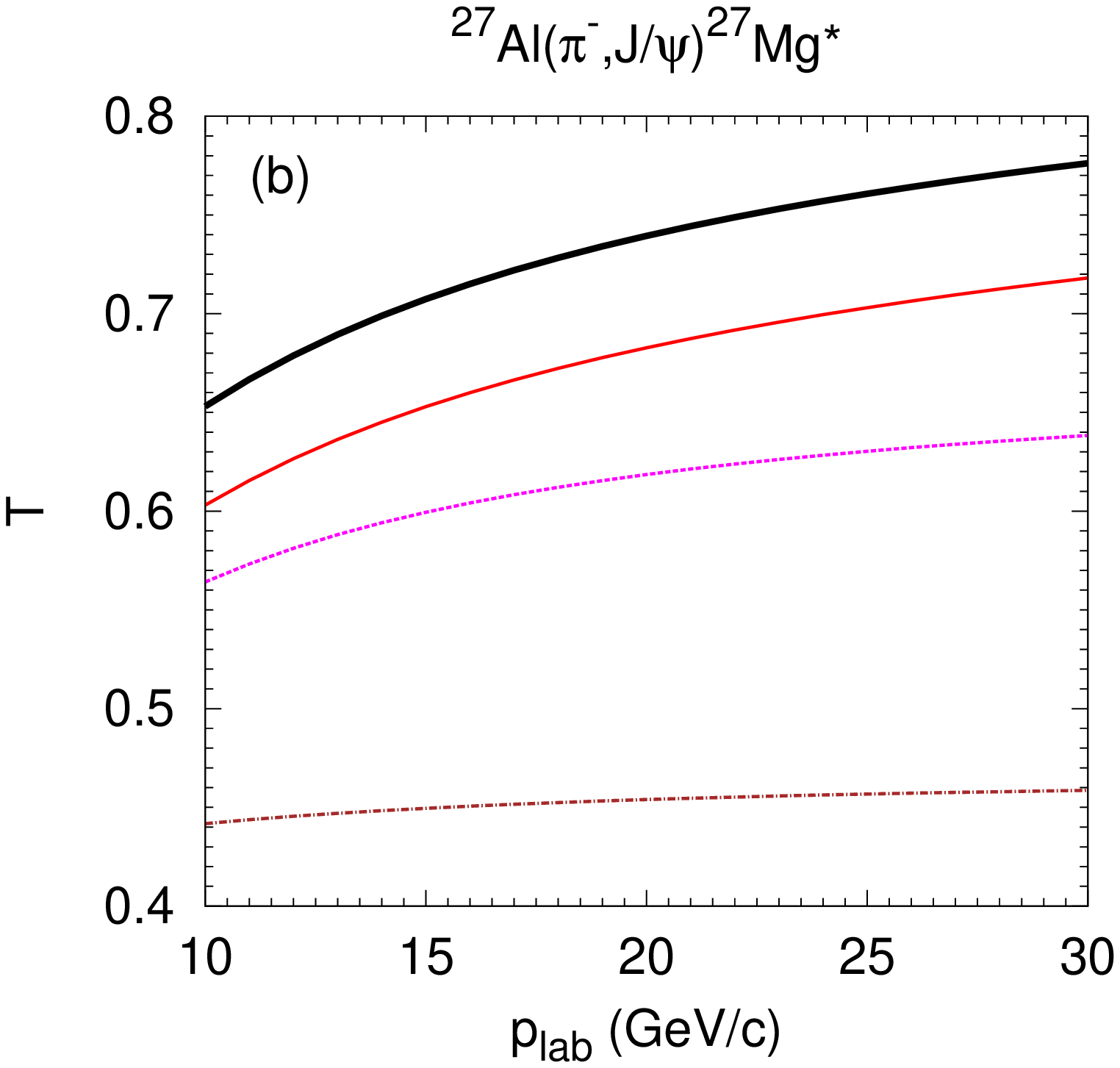} \\
\includegraphics[scale = 0.4]{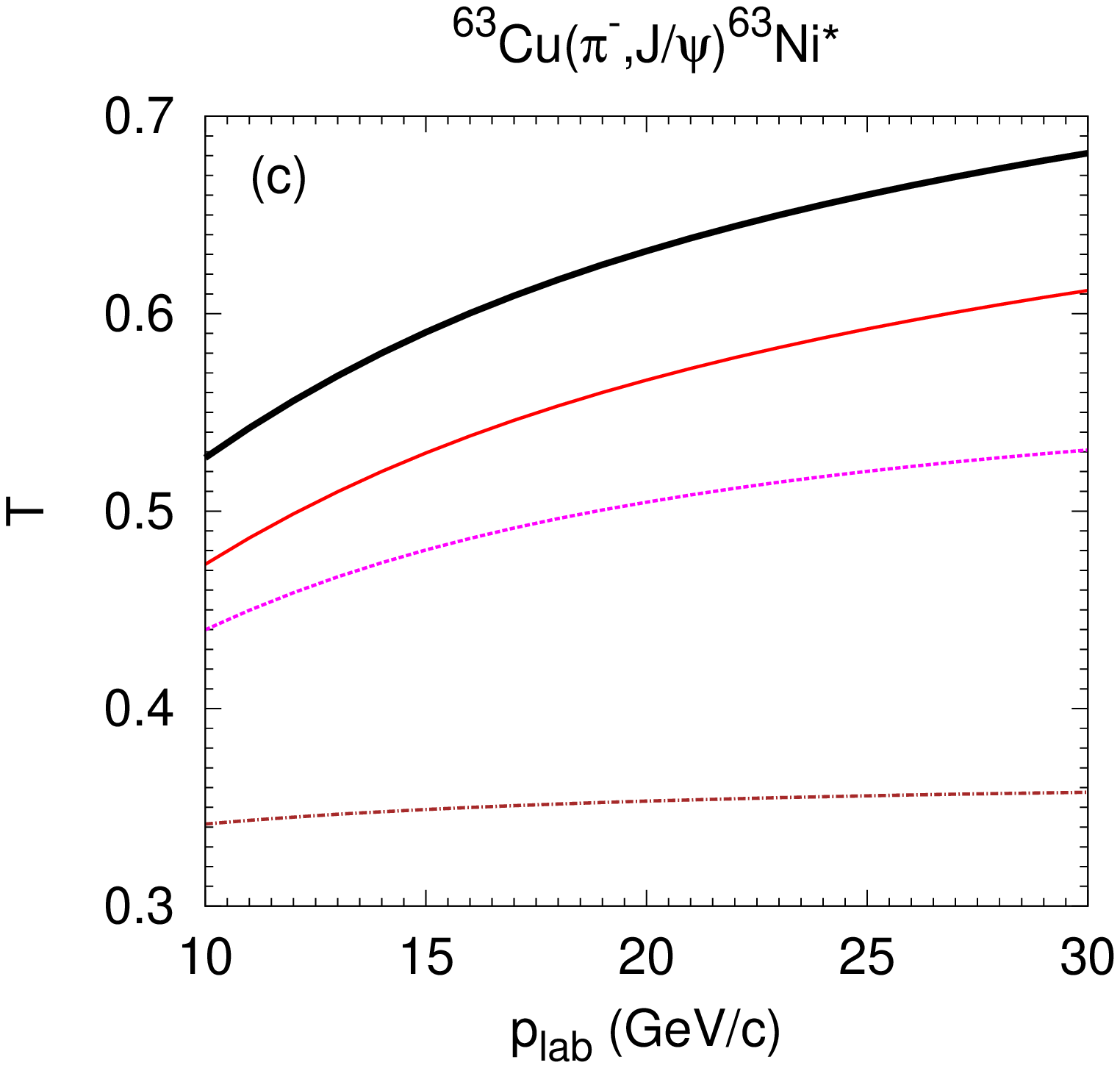} &
\includegraphics[scale = 0.4]{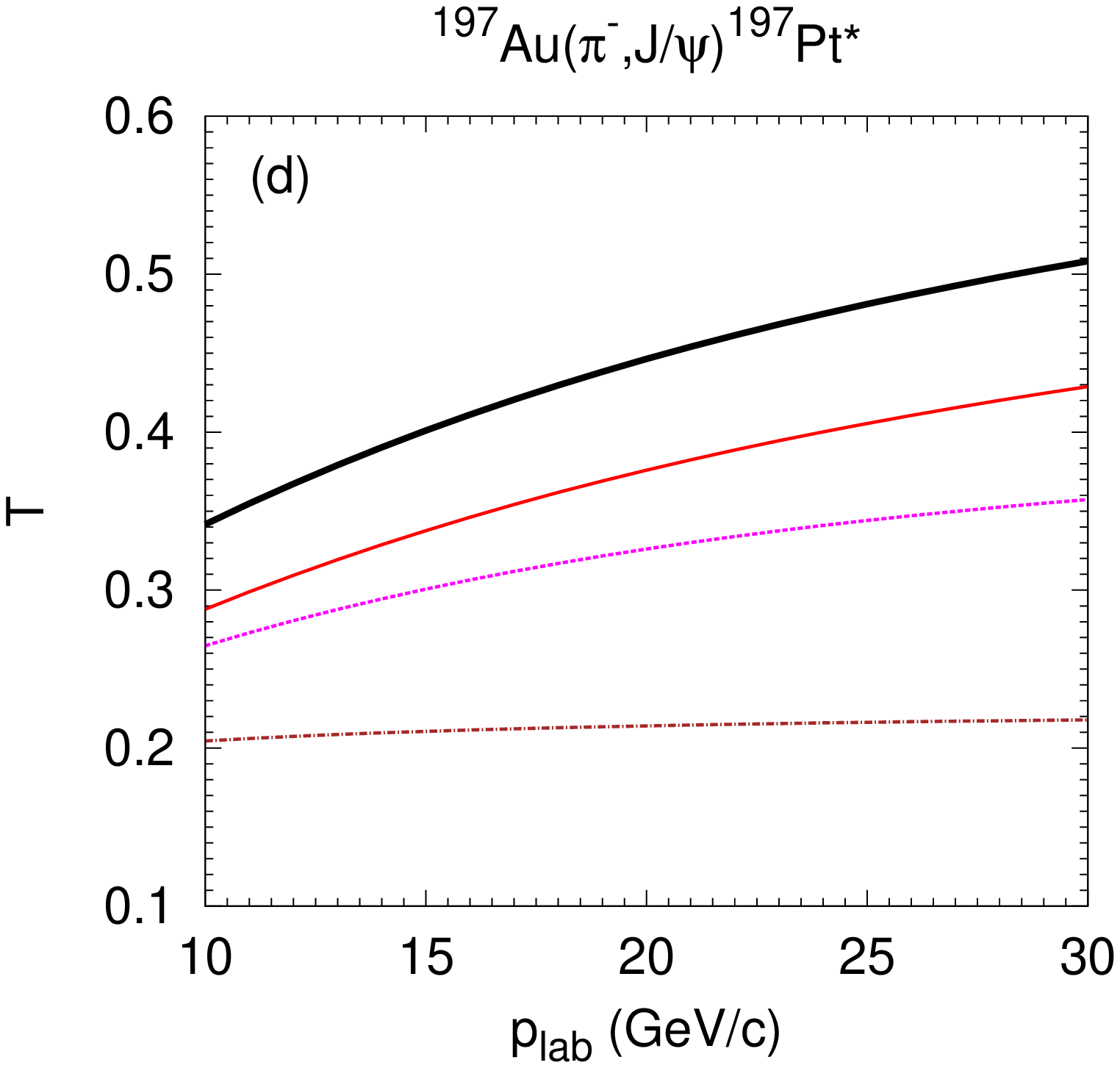} \\
\end{tabular}
\caption{\label{fig:T_pi_Jpsi} (color online) Transparency vs beam momentum for the $(\pi^-,J/\psi)$ reaction 
on $^{12}$C, $^{27}$Al, $^{63}$Cu, and $^{197}$Au targets (panels (a), (b), (c), and (d), respectively). 
The calculations are done within the Glauber and quantum diffusion models by setting the total $J/\psi N$ cross section 
equal to 4 and 6 mb as indicated. The curves marked with 'CT'  are calculated taking into account color transparency effects
both for incoming pion and outgoing $J/\psi$. The curves marked with 'CT pion' are calculated with the color transparency
effect for the pion only. See also text.}
\end{figure}
In the case of $J/\psi$ production in $\pi^-$-induced reactions, the transparency is displayed in Fig.~\ref{fig:T_pi_Jpsi}.
The behaviour of the $J/\psi$ transparency with beam momentum is close to that of dilepton transparency 
(cf. Fig.~\ref{fig:T_4gev2}), since the $J/\psi N$ cross section is small. The effect of the CT is 5-10 \% stronger 
for the $J/\psi$ transparency than for the dilepton transparency as in the former case the CT influences both the incoming 
and outgoing particles.
To separate the effects of the CT on the pion and on $J/\psi$, we also show in Fig.~\ref{fig:T_pi_Jpsi} the calculation
with the cross section in the hard interaction point set equal to the 'normal' $J/\psi N$ cross section (i.e. to 6 mb).
This is equivalent to keeping the CT effect for the pion only. Dropping the CT effect for the outgoing charmonium
significantly reduces the transparency, in particular for high beam momenta and for the light nuclei.
However, for the $^{197}$Au nucleus the reduction of the transparency at the beam momenta $10-15$ GeV/c is quite moderate 
and allows to disentangle various $J/\psi N$ cross sections.
\begin{figure}
\includegraphics[scale = 0.6]{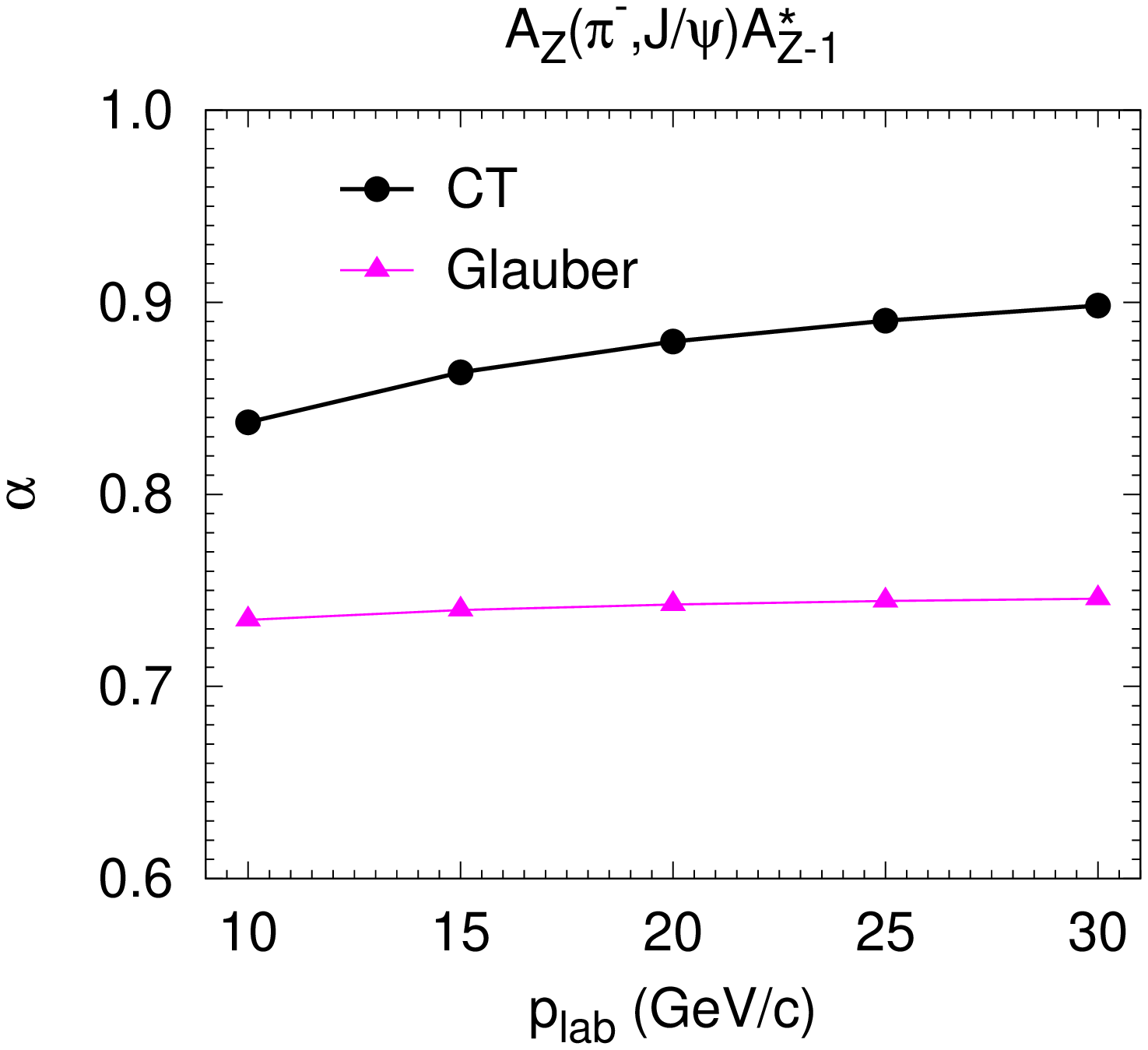}
\caption{\label{fig:alpha_pi_Jpsi} (color online) Beam momentum dependence of the parameter $\alpha$ of the
power law fit (\ref{T_vs_A}) of the transparency for the $A_Z(\pi^-,J/\psi)A^*_{Z-1}$ reaction
calculated within the quantum diffusion model (line with solid circles) and within the Glauber model 
(line with solid triangles).}
\end{figure}
The power law exponent $\alpha$ for the $J/\psi$ production reaction is shown in Fig.~\ref{fig:alpha_pi_Jpsi}. 
In general, the $\alpha(p_{\rm lab})$ dependencies are similar for the $J/\psi$ and dilepton production
(cf. Fig.~\ref{fig:alpha}).

\begin{figure}
\includegraphics[scale = 0.6]{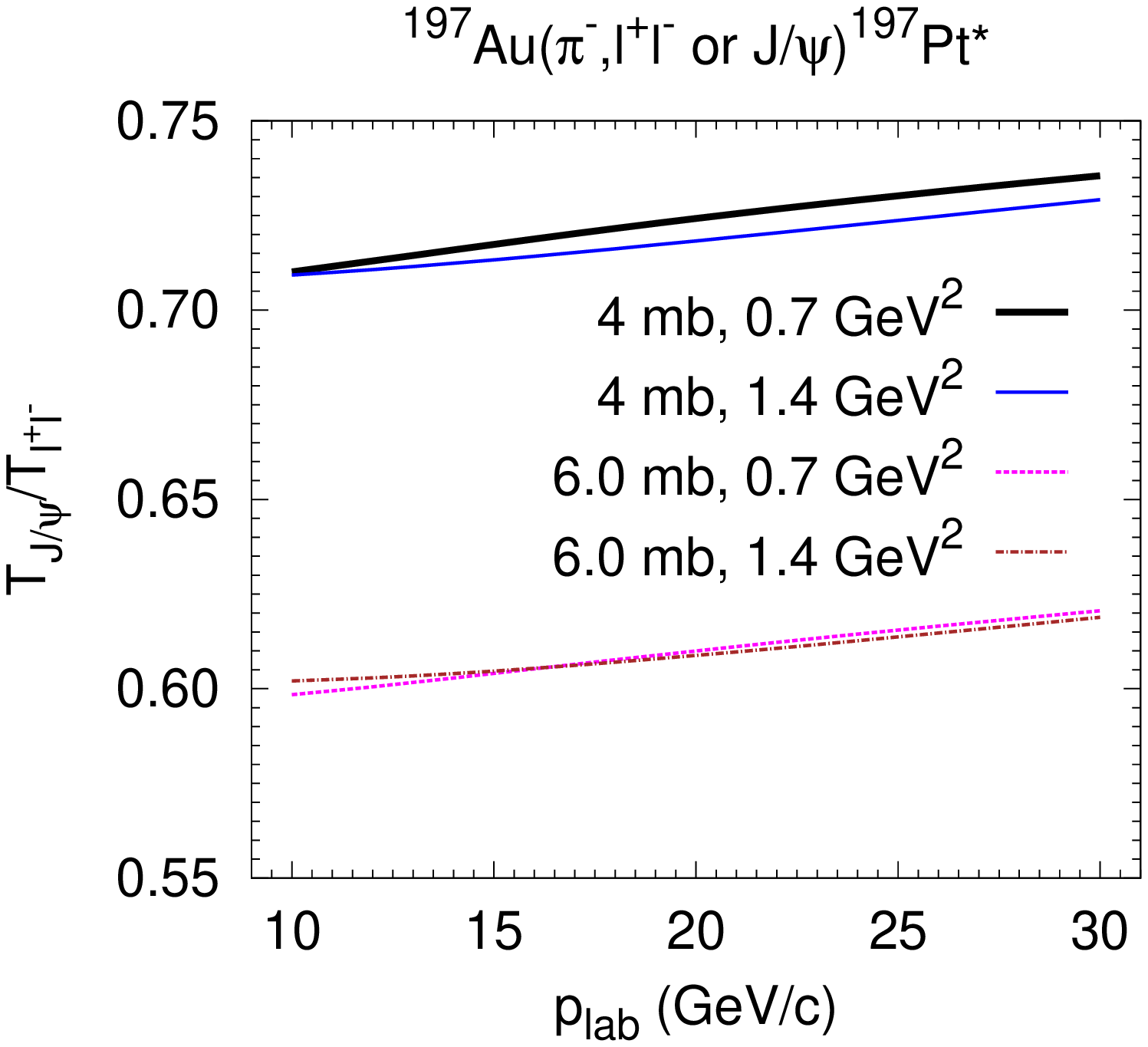}
\caption{\label{fig:ratio_Au197_mJpsi} (color online) The ratio of transparencies for the $J/\psi$ and $l^+l^-$ outgoing
channels in $\pi^-$ interactions with $^{197}$Au vs beam momentum. 
Calculations are done within the quantum diffusion model for the different values of total $J/\psi N$ cross section
and the $\Delta M^2$ parameter in the pion coherence length as indicated.}
\end{figure} 
In the experiment, one could perform the two sets of measurements of the $l^+l^-$ transparency over a range of pion beam momenta.
In the first measurement, one can fix $M_{l^+l^-}=M_{J/\psi}$. In the second measurement, one can fix the dilepton mass slightly
above the $J/\psi$ mass (fixing it below $M_{J/\psi}$ is impossible because of the contributions of radiative corrections).
Since the width of the $J/\psi$ is small, $93$ keV, in the second measurement the $J/\psi$ will be excluded in the $l^+l^-$ channel. 
Thus, the transparency should strongly vary between such two measurements.
This is illustrated in Fig.~\ref{fig:ratio_Au197_mJpsi}, which shows the ratio of the transparencies for the $J/\psi$ and $l^+l^-$
outgoing channels, where the latter is considered as a background at $M_{l^+l^-} = M_{J/\psi}$. The ratio is insensitive to the
possible uncertainties of the pion coherence length, as displayed in Fig.~\ref{fig:ratio_Au197_mJpsi}.   
Note that the degree of pion squeezing in the hard interaction point is not the same in calculations with outgoing $J/\psi$
and $l^+l^-$ (see Sec. \ref{model}). However, this difference has only minor influence on the ratio of transparencies, as
the free $\pi N$ cross section is in any case much larger than the cross section of interaction of the squeezed 
$q \bar q$ configuration with the nucleon.
Thus, the ratio of the transparencies can be used to study the genuine $J/\psi N$ cross section with an accuracy 
similar to that of $J/\psi$ photoproduction at SLAC \cite{Anderson:1976hi,Larionov:2013axa}.     

Another interesting channel is the exclusive production of $\chi$-mesons. This cross section may be larger than for $J/\psi$'s 
since in the $\chi$ case the two gluon exchange is allowed while for $J/\psi$ three gluons are necessary. 
Transparency for $\chi$ production is likely to be slightly smaller than in the $J/\psi$ case since
the $c \bar c$ pair has to squeeze to $r_\chi$ which is larger than $r_{J/\psi}$ and $\sigma_{J/\psi N} < \sigma_{\chi N}$.

\section{Conclusions}
\label{concl}

In this work, we considered the observable consequences of a possible transverse squeezing of the incoming pion in the semiexclusive
$A(\pi^-,l^+l^-)$ reaction with large invariant mass of the dilepton pair.
This corresponds to the kinematic range complementary to the one probed in DIS where the exchange photon is space-like. 
Assuming that the rate of squeezing is similar to the one observed in DIS pion production (that is given by the  $1/M_{l^+l^-}$ scale)
we predict large ($\ge 50\%$) color transparency effects in the explored kinematic regime. These effects are accessible in experiments 
with pion beams with momenta  around 10 -- 20 GeV/c. The increase of the nuclear transparency due to the color transparency is more prominent 
for heavy targets than for light targets.

By a similar mechanism, squeezing of the incoming pion is expected for the semiexclusive $A(\pi^-,J/\psi)$ reaction 
(as well as the production of $\chi$-mesons)
which is driven by the small $(\mbox{charmonium mass})^{-1}$ scale. Thus, the color transparency signal should be observable in this case too. 
Taking into account the uncertainty of the degree of squeezing of the charmonium in the hard interaction point, 
the genuine $J/\psi N$ cross section can be determined with an accuracy better than $\sim 20\%$ from the ratio of transparencies for
the $J/\psi$ and pure dilepton outgoing channels. 

We have also given predictions for the nuclear transparency in the case of $\pi^-$-induced $\rho^0$ production on nuclei using the Glauber model.
The comparison with the $\rho$ meson photoproduction would allow to check whether similar configurations in the $\rho$ meson dominate 
for the interaction in the vacuum and non-vacuum channels.

The effects discussed in this work can be experimentally studied with pionic secondary beam which will be
available at the J-PARC and, possibly, at GSI/FAIR.

\begin{acknowledgments}
AL and MB acknowledge financial support of HIC for FAIR within the framework of
the Hessian LOEWE program. MS's work is supported by US Department of Energy grants 
under contract DE-FG02-93ER40771.
\end{acknowledgments}

\bibliography{pionCT}

\end{document}